\documentclass[twocolumn, superscriptaddress, aip, preprintnumbers, showpacs, floatfix, amsmath,amssymb,showkeys,10pt]{revtex4-1}

\usepackage{graphicx}
\usepackage{hyperref}
\hypersetup{backref,pdfpagemode=FullScreen,colorlinks=true,allcolors=blue}

\graphicspath{{./Figures/}}
\usepackage{float}

\usepackage{multirow}
\usepackage{tabularx}
\usepackage[table]{xcolor}

\usepackage{tikz}
\usetikzlibrary{shapes,arrows,calc}
\usepackage{makecell}

\usepackage[caption=false]{subfig}

\everymath{\displaystyle}

\usepackage{tikz}
\usetikzlibrary{calc,arrows,shapes,positioning}

\tikzstyle{decision} = [rectangle, draw=pp, fill=pp, 
    text width=10em, text centered, rounded corners, minimum height=3em,line width=1mm]
\tikzstyle{block} = [rectangle, draw=qe, fill=qe, 
    text width=10em, text centered, rounded corners, minimum height=3em,line width=1mm]

\tikzstyle{line} = [draw=arsenic, -latex', line width=0.1cm]

\tikzstyle{cloud} = [ellipse, draw=ym, fill=ym, 
    text width=7em, text centered, rounded corners, minimum height=4em,line width=1mm]

\usepackage{mathtools}
\usepackage[most]{tcolorbox}
\usepackage{bbm}
\usepackage{multirow}

\usepackage{bbm}
\usepackage{calc}  
\usepackage{enumitem}  
\usepackage{setspace}
\usepackage{float}
\usepackage{indentfirst}
\usepackage{fixltx2e}

\definecolor{pp}{RGB}{200,200,200}
\definecolor{ym}{RGB}{255, 102, 163}
\definecolor{qe}{RGB}{255, 179, 102}

\definecolor{sangria}{rgb}{0.57, 0.0, 0.04}
\definecolor{arsenic}{rgb}{0.23, 0.27, 0.29}

\definecolor{prussianblue}{rgb}{0.0, 0.19, 0.33}
\definecolor{phthalogreen}{rgb}{0.07, 0.21, 0.14}
\definecolor{charcoal}{rgb}{0.21, 0.27, 0.31}

\tcbset{highlight math style={enhanced,
  colframe=sangria,colback=arsenic!15,arc=0pt,boxrule=0.5mm}}

\tikzstyle{mybox} = [draw=arsenic, fill=arsenic!10, very thick,
    rectangle, rounded corners, inner sep=10pt, inner ysep=15pt]

\tikzstyle{figbox} = [draw=white, fill=white, very thick,
    rectangle, inner sep=30pt, inner ysep=5pt]

\tikzstyle{fancyarsenic} =[draw=arsenic, fill=arsenic!10, very thick,
    rectangle, rounded corners, text=arsenic, inner sep=10pt, inner ysep=8pt]    
  
\tikzstyle{fancyred} =[draw=arsenic, fill=arsenic!10, very thick,
    rectangle, rounded corners, text=sangria, inner sep=10pt, inner ysep=8pt]  

\tikzstyle{fancyblue} =[draw=arsenic, fill=arsenic!10, very thick,
    rectangle, rounded corners, text=prussianblue, inner sep=10pt, inner ysep=8pt]  
    
\tikzstyle{fancygreen} =[draw=arsenic, fill=arsenic!10, very thick,
    rectangle, rounded corners, text=phthalogreen, inner sep=10pt, inner ysep=8pt]      

\def\alloy#1#2#3{Au$_\mathrm{#1}$Ag$_\mathrm{#2}$Cu$_\mathrm{#3}$ }
\def\ep{\mathcal{E}}
\def\td#1{$_\mathrm{#1}$}
\def\tu#1{$^\mathrm{#1}$}
\def\der#1{d\mathbf{#1}\;}
\def\im{\mathit{i}}
\def\gw{G$_0$W$_0$ }
\def\gwr{G$_0$W$_0$+RPA }
\def\Re#1{\mathfrak{Re}\left[#1\right]\;}
\def\Im#1{\mathfrak{Im}\left[#1\right]\;}

\begin{document}

\title{Plasmonic performance of Au$_\mathbf{x}$Ag$_\mathbf{y}$Cu$_\mathbf{1-x-y}$ alloys from many-body perturbation theory}

\author{Okan K. Orhan}
 %\altaffiliation[Also at ]{School of Physics, Trinity College Dublin.}%Lines break automatically or can be forced with \\
\author{David D. O'Regan}%
 %\email{Second.Author@institution.edu}
\affiliation{%
School of Physics, Trinity College Dublin, Dublin 2, Ireland\\
}

\date{\today}
\begin{abstract} 
We present a detailed appraisal of the optical and plasmonic properties
of ordered alloys of the form Au$_{x}$Ag$_{y}$Cu$_{1-x-y}$, as predicted by means of first-principles many-body perturbation theory augmented by a semi-empirical Drude-Lorentz model.
In  benchmark simulations on elemental Au, Ag, and Cu, 
we find that the random-phase approximation (RPA) 
fails to accurately describe inter-band transitions when it is built
upon semi-local approximate Kohn-Sham density-functional theory (KS-DFT) band-structures.
We show that non-local electronic 
exchange-correlation interactions sufficient 
to correct this, particularly for the 
fully-filled, relatively narrow $d$-bands that which contribute strongly throughout the low-energy spectral range ($0-6$~eV), may be modelled
very expediently using band-stretching operators that imitate
the effect of a  perturbative \gw self-energy correction
incorporating quasiparticle mass renormalization.
We thereby establish a convenient  work-flow for carrying out approximated
\gwr spectroscopic calculations on alloys and,   in particular here, we have considered alloy concentrations
down to $12.5~$\% in \alloy{x}{y}{1-x-y}, including all possible crystallographic orderings
of face-centred cubic (FCC) type.
We develop a pragmatic procedure for calculating the Drude 
plasmon frequency from first principles, including self-energy effects, 
as well as a semi-empirical scheme for
interpolating the  plasmon inverse lifetimes between stoichiometries.
A distinctive M-shaped profile is observed in both quantities
for binary alloys, in qualitative agreement with previous experimental findings.
A range of optical and plasmonic figures of merit are discussed, 
and plotted for ordered \alloy{x}{y}{1-x-y} at
three representative solid-state laser wavelengths.
On this basis, 
we predict that certain compositions may offer improved
performance over elemental Au for particular application types.
We predict that while the loss functions for both bulk and 
surface plasmons are typically diminished in strength
through binary alloying, certain stoichiometric ratios 
may exhibit higher-quality (longer-lived) 
localized surface-plasmons (LSP) and surface-plasmon polaritons (SPP),  at technologically-relevant wavelengths, than those
in elemental Au.

\end{abstract}


\maketitle

%\submitto{\JPCM}

\section{Introduction}
Noble metals  and their alloys are compelling materials for  opto-electronic applications~\cite{black2000magnetic,doi:10.1021/jp062536y,doi:10.1021/ar7002804,zou2014recording,Medici2015329,sato2009heat} due to their strong plasmonic and optical response throughout the infrared-visible-ultraviolet spectral range.
The tailoring of plasmonic and optical properties of  metals via alloying  is currently attracting interest due to a high demand for  novel and more efficient nano-materials for  opto-electronic applications~\cite{liu2008cu,VALODKAR2011384,ADOM:ADOM201500446}.
Nano-structures of  pure Au, Ag and Cu are  used  in diverse opto-electronic applications, due to their good chemical and mechanical stabilities, as well as their strong optical response in the low-energy spectral range.
 Hence, their alloys are naturally expected to be promising candidates for efficient opto-electronic applications.
Spectroscopic measurements on alloys are mostly performed on their thin-film surfaces~\cite{PhysRev.132.2062,PhysRev.140.A2105,stahl1969optical,doi:10.1143/JPSJ.30.399,PhysRevB.6.1209,
RIVORY1976345,Nishijima:12,Nishijima2014,doi:10.1021/acsphotonics.5b00586,hashimoto2016ag}, which are highly dependent on the alloying technique used~\cite{Pena-Rodriguez:14}.
As a result, it is difficult to find consensus 
within the literature even on basic quantities such as the plasma frequency of an elemental metal.

Systematic first-principles studies of such alloys may, therefore,  offer fundamental insights into the microscopic effects of alloying on optical properties, and potentially thereby even guide the tailoring of optical and plasmonic response for designated applications.
This work is an exploratory investigation into the viability of such an approach using state-of-the-art theory.
Specifically, in this article, we present a detailed 
 investigation into the capabilities and  limitations of  contemporary theoretical spectroscopy for noble metals, and  the development and testing of a set of computationally light techniques for studying the spectra of noble metals and their ordered alloys within the linear-response regime. 
Taking advantage of the resulting high-throughput-compatible approach, we provide various \emph{figures of merit} for 
comparing the plasmonic performance of these alloys, from which the optimal stroichiometry for a given optical or plasmonic characteristic, at a given driving frequency, may be estimated.

\section{Theoretical Methodology}\label{sec:s1}
In the study and simulation of solid-state optical and energy-loss spectra, the macroscopic dielectric function 
$\ep(\omega)$
is the central function due to its well-established connection to measured observables.
In the low-energy spectral range, where the phenomena contributing to optical spectra are almost  entirely electronic, the macroscopic dielectric function of a metallic system is constituted by two terms, explicitly by 
\begin{align}\label{eq:s2e1}
\ep(\omega)=\ep^{\mathrm{inter}}(\omega)+\ep^{\mathrm{intra}}(\omega),
\end{align} 
where $\ep^{\mathrm{inter}}(\omega)$ and $\ep^{\mathrm{intra}}(\omega)$ result from screening effects due to inter-band transitions and  intra-band transitions  (giving rise to the Drude plasmon), respectively. 
When a solid is  simulated  as an infinite object, electronic transitions are envisaged as occurring  simultaneously throughout the material, 
at a given energy of electro-magnetic (EM) radiation.
Observable spectra are spatial averages of these transitions,  whereas
 electronic transitions are microscopic events.
Hence, an averaging process is required to connect  microscopic quantities to 
macroscopic observables. 
Specifically, an averaging via the inverse dielectric function is the appropriate route to obtain the macroscopic dielectric function in the optical (vanishing momentum transfer) limit~\cite{PhysRev.126.413,PhysRev.129.62}, and this is given, 
in terms of the \emph{microscopic}  dielectric matrix
$\varepsilon$, by
\begin{align}\label{eq:s2e2}
\ep(\omega)=\frac{1}{\lim_{\mathbf{q}\rightarrow 0}\left[\varepsilon^{-1}_{\mathbf{G},\mathbf{G}'}(\mathbf{q},\omega)\right]_{\mathbf{G},\mathbf{G}'=0}}.
\end{align}
As the momentum $\mathbf{q}$ transferred from the incoming photon is assumed to be negligible, only vertical excitations at each point in reciprocal space are considered here.
Notwithstanding,  non-local screening `local field' effects are explicitly incorporated by means of  Eq.~\ref{eq:s2e2}.
Where such effects can be safely neglected,  
we may used the relatively simple, 
conveniently inversion-free formula
\begin{align}
\label{neglect-nlf}
\ep(\omega)\approx\lim_{\mathbf{q}\rightarrow 0}\left[\varepsilon_{\mathbf{G},\mathbf{G}'}(\mathbf{q},\omega)\right]_{\mathbf{G},\mathbf{G}'=0}.
\end{align}

\subsection{Inter-band transitions}

Inter-band transitions are  electronic transitions between 
the valence (occupied) 
and conduction (unoccupied) bands.
Within the linear-response regime that typically
holds for photon energies in the IR-vis-UV range, the 
 inverse microscopic dielectric function is 
 given, in reciprocal space, by the expression 
\begin{align}\label{eq:s2e3}
\varepsilon^{-1}_{\mathbf{G},\mathbf{G}'}(\mathbf{q},\omega)=
\delta_{\mathbf{G},\mathbf{G}'} + v_{\mathbf{G},\mathbf{G}'}(\mathbf{q})\chi_{\mathbf{G},\mathbf{G}'}(\mathbf{q},\omega),
\end{align}
where the bare Coulomb interaction  takes the  form
\begin{align}\label{eq:s2e4}
v_{\mathbf{G},\mathbf{G}'}(\mathbf{q})=\frac{4\pi}{|\mathbf{q}+\mathbf{G}||\mathbf{q}+\mathbf{G}'|}.
\end{align}
In order to calculate 
$\varepsilon_{\mathbf{G},\mathbf{G}'}(\mathbf{q},\omega)$,
the  non-interacting random-phase approximation (RPA)
 (Fermi's Golden Rule)~\cite{dirac1927quantum} 
 linear-response function for  inter-band excitations is  first calculated, in terms of  independent-particle wave-functions
 $|\psi_{v,\mathbf{k}}\rangle$
 with occupancies $f_{v,\mathbf{k}}$, as~\cite{PhysRev.126.413,PhysRev.129.62},
\begin{align}\label{eq:s2e5}
 \chi^0_{\mathbf{G}\mathbf{G}'}(\mathbf{q},\omega) ={}&  2\sum_{c,v,\mathbf{k}} 
(f_{v,\mathbf{k}}-f_{c,\mathbf{k}-\mathbf{q}}) 
\\   \nonumber
{}& \times
 \frac{
| \langle\psi_{c,\mathbf{k}-\mathbf{q}}|e^{\im (\mathbf{q}+\mathbf{G})\cdot\mathbf{r}}|\psi_{v,\mathbf{k}}\rangle |^2
%
%\langle \psi_{v,\mathbf{k}}|e^{-\im (\mathbf{q}+\mathbf{G}')
%\cdot\mathbf{r}}|\psi_{c,\mathbf{k}-\mathbf{q}}\rangle
}{\omega-\omega_{cv,\mathbf{k}\mathbf{q}}+\im\Gamma},
\end{align}
where $v$ $(c)$ indicates valence (conduction) states and
 $\omega_{cv,\mathbf{k}\mathbf{q}}=(E_{c,\mathbf{k}-\mathbf{q}}-E_{v,\mathbf{k}})$ is a difference between
single-particle energy eigenvalues (using atomic units).
Here, $\Gamma$ is a small, positive-valued  
Lorentzian broadening factor, $\mathbf{q}$ is the transferred momentum vector, which lies in the first Brillouin zone, and the factor of $2$ pre-supposes and accounts for spin degeneracy.
Following this, the \emph{interacting} RPA~\cite{PhysRev.82.625,PhysRev.85.338,PhysRev.92.609,RevModPhys.36.844} response function 
$\boldsymbol{\chi}$
is given
by~\cite{PhysRev.82.625,PhysRev.85.338,PhysRev.92.609,RevModPhys.36.844} 
\begin{align}\label{eq:s2e6}
\hspace{-0.2cm} \chi_{\mathbf{G}\mathbf{G}'}(\mathbf{q},\omega)=
\chi^0_{\mathbf{G}\mathbf{G}'}(\mathbf{q},\omega)\left(1+v_{\mathbf{G},\mathbf{G}'}(\mathbf{q})\chi_{\mathbf{G}\mathbf{G}'}(\mathbf{q},\omega)\right),
\end{align}
which is a Dyson equation. Eq.~\eqref{eq:s2e6}
may be rearranged into the compact form, involving matrix inversion, of 
\begin{align}\label{eq:s2e7}
\boldsymbol{\chi}^{-1}
(\mathbf{q},\omega)=\boldsymbol{\chi}_0^{-1}
(\mathbf{q},\omega)-\mathbf{v}(\mathbf{q}).
\end{align}
for the interacting response function.
Alternatively, the more approximate 
independent-particle RPA dielectric
function may be calculated as
\begin{align}\label{eq:s2e8}
\varepsilon^{0}_{\mathbf{G},\mathbf{G}'}(\mathbf{q},\omega)=\delta_{\mathbf{G},\mathbf{G}'}-v_{\mathbf{G},\mathbf{G}'}(\mathbf{q})\chi^0_{\mathbf{G}\mathbf{G}'}(\mathbf{q},\omega).
\end{align}
When $\varepsilon \approx
\varepsilon^0$ is invoked it is conventional 
to also  neglect  local-field effects, by 
means of Eq.~\ref{neglect-nlf}.

\subsection{Intra-band transitions}

In addition to inter-band transitions, the promotion of electrons to higher energies within the \emph{same} band, at finite temperatures,
contributes very significantly 
to the macroscopic dielectric function of metallic solids
within the low-energy regime.
This is the intra-band transition effect, which gives rise to the prominent Drude plasmon divergence in the optical absorption spectrum in the static limit.
This Drude plasmon can be thought of semi-classically as the collective oscillation of electrons at the Fermi level, in phase with the longitudinal part of the driving EM radiation.
The Drude plasmon typically occurs at $\sim10$~eV in elemental late transition metals, and it can be  excited, e.g., by energy loss of incident electrons with  kinetic energies in the $1-20$~keV range, or by using lasers tuned to the plasmon wavelength.

The direct simulation of a well-converged 
Drude plasmon frequency starting a set of
single-particle electronic states is  computationally demanding, indeed extremely so due to the requirement for  dense Brillouin-zone sampling at and around the Fermi surface, for example up to $\sim16000$ grid points in Ref.~\onlinecite{marini2001optical}.
This procedure is not commonly followed, as the resulting 
intra-band dielectric function remains 
excessively  sensitive to the 
difficult-to-estimate excitation damping (or lifetime)
factor that must be imposed.
The Drude plasmon lifetime is limited by
a multitude of physical processes, in reality,  including scattering by phonons, defects, and grain boundaries; electron-electron (including electron-plasmon and plasmon-plasmon) scattering giving rise to decay, and quantum thermodynamic  effects.

More commonly,  the Drude plasmon is discussed in terms of a classical model for  free electrons oscillating under the influence of external electric field, namely the Drude-Lorentz model~\cite{ANDP:ANDP19003060312,ANDP:ANDP19003081102,lorentz1909theory,fox2002optical}.
The Drude plasmon contribution to  the macroscopic dielectric function becomes 
\begin{align}\label{eq:s2e9}
\mathcal{E}^{\mathrm{intra}}(\omega)=
\varepsilon_{\infty}-\frac{\omega_\mathrm{p}^2}{\omega^2-\im \eta_\mathrm{p} \omega},
\end{align}
where $\omega_\mathrm{p}$, $\eta_\mathrm{p}$, and 
$\varepsilon_{\infty}$ are the Drude plasmon energy
(not to be confused with the actual net plasma frequency of the metal -- where $\Re{\mathcal{E}(\omega)}\approx 0$~eV),  the phenomenological inverse life-time, and the electric permittivity in the infinite-frequency limit,  respectively. 
The set $\{\omega_\mathrm{p},\eta_\mathrm{p},\varepsilon_{\infty}\}$ comprise the `Drude parameters'.
Experimentally, it is standard practice is to perform  optical (e.g., $n$ and $k$) measurements within the infrared and the far-infrared spectral range (corresponding to $\omega \approx 0-2$~eV), and to determine these parameters by fitting to the Drude-Lorentz model~\cite{Ordal:85,doi:10.1021/j100287a028,PhysRevB.65.165432,doi:10.1021/nl050062t,Grady2004167,PhysRevB.65.125415,doi:10.1021/jp810808h}.
%
%Unfortunately, determining  the Drude parameters is not possible solely from the electronic bands directly;  however a step-by-step semi-empirical approach can be used to approximate them.

Inspired by this, here, we start from the Drude plasmon energy in Eq.~\eqref{eq:s2e9}, which can be expressed as~\cite{ANDP:ANDP19003060312,ANDP:ANDP19003081102,PhysRevB.16.5277}
\begin{align}\label{eq:s2e10}
\omega_\mathrm{p}^2=\frac{4\pi N(E_\mathrm{F})}{m_{\mathrm{eff}}},  
\end{align}
for a uniform non-interacting electron gas.
In this, $N(E_\mathrm{F})$ is  the density of states (DOS) at  the Fermi level and $m_{\mathrm{eff}}$ is the 
electron effective mass.
In practice, for real metals, this effective mass is also evaluated
at the Fermi level and, if we further assume that
the metallic bands have a parabolic  dispersion
normal to the Fermi surface~\cite{doi:10.1080/14786435808237011},
we may write %
\begin{align}\label{eq:s2e11}
m_{\mathrm{eff}}^{-1} &{}\approx  \frac{1}{3} \langle v^2(E_\mathrm{F})\rangle 
\nonumber \\ &{}=
\frac{1}{3}
 \Big(\sum_i\int_{S_\mathrm{F_i}}\der{k} v^2_i(\mathbf{k}) 
 \Big)
 \Big( \sum_j \int_{S_\mathrm{F_j}}\der{k'} \Big)^{-1},  \nonumber \\
\mbox{if} \quad
v^2_i(\mathbf{k})&{}=
\left|\frac{\partial E_{i,\mathbf{k}}}{\partial \mathbf{k}}\right|^2.
\end{align}
Here,  $S_\mathrm{F_i}$ signifies the Fermi surface
of the $i^\textrm{th}$ metallic band, and the factor
of $1/3$ results from the squared Fermi velocity 
being averaged (rather than summed) over
Cartesian directions.
Succinctly, the Drude plasmon energy  can 
thus be approximated
 within a non-interacting, uniform-gas theory,  
 simply and efficiently 
 as~\cite{doi:10.1080/14786435808237011}
\begin{align}\label{eq:s2e12}
\omega_\mathrm{p}^2=\frac{4\pi}{3} N(E_\mathrm{F})\langle v^2(E_\mathrm{F}) \rangle.
\end{align}

As previously mentioed,  the routine 
direct calculation of experimentally-relevant
Drude plasmon lifetimes for is currently 
beyond the scope of state-of-the-art 
electronic structure simulation methodology.
Electron-phonon and electron-impurity scattering 
typically dominantly contribute to the limiting 
DC conductivity  $\sigma_0$, as
compared  to the more accessible electron-electron scattering processes~\cite{PhysRevB.16.5277}. 
In order to circumvent this issue, we have developed 
a semi-empirical scheme based upon the Drude-Lorentz
model, in which the scattering rate 
$\eta_\mathrm{p}$
is inversely proportional to the
DC conductivity and to the  effective mass of carriers, 
but proportional to their concentration.
Noting a very plausibile linear dependence between 
$\sigma_0$ and $\sqrt{\langle v^2(E_F)\rangle}$, 
we express the  scattering rate 
(where the first equality is standard Drude-Lorentz) as 
\begin{align}\label{eq:s2e13}
\eta_\mathrm{p}=\frac{ N(E_\mathrm{F})}{\sigma_0 m_{\mathrm{eff}}}\approx c_{\eta}N(E_\mathrm{F}) \langle v^2(E_\mathrm{F})\rangle^{1/2},
\end{align}
where $c_{\eta}$ is our scaling coefficient to be determined.

Next,  separating the real and imaginary part of $\ep_{\mathrm{intra}}(\omega)$ from Eq.~\eqref{eq:s2e9}, we arrive at
\begin{align}\label{eq:s2e14}
\Re{\mathcal{E}^{\mathrm{intra}}}=\mathcal{E}^{\mathrm{intra}}_1(\omega,\omega_\mathrm{p},\eta_\mathrm{p},\varepsilon_{\infty})&=\varepsilon_{\infty}-\frac{\omega_\mathrm{p}^2}{\omega^2+\eta_\mathrm{p}^2},\nonumber \\
\Im{\mathcal{E}^{\mathrm{intra}}}=\mathcal{E}^{\mathrm{intra}}_2(\omega,\omega_\mathrm{p},\eta_\mathrm{p})&=\frac{\eta_\mathrm{p}\omega_\mathrm{p}^2}{\omega^3+\eta_\mathrm{p}^2\omega},
\end{align}
and note that, given the first-principles $\omega_\mathrm{p}$,  the imaginary part is  
parametrized only by $\eta_\mathrm{p}$. 
Thus, in practice, we first determine the scaling factor
$c_{\eta}$
 for $\eta_\mathrm{p}$ in Eq.~\eqref{eq:s2e13} by 
 least-squares fitting against
 $\mathcal{E}^{\mathrm{intra}}_2$ from an
 appropriate experimental spectrum within 
 the near infra-red spectral range 
 (we find that $\omega \approx 0.8-1.2$~eV
 is very effective for noble metals -- but we emphasize
 that the choice of regression domain does matter).
Secondly, we determine $\varepsilon_{\infty}$ by fitting $\mathcal{E}^{\mathrm{intra}}_1$ to the same 
experimental spectrum, using the fixed values for 
$\omega_\mathrm{p}$ and $\eta_\mathrm{p}$ obtained in the previous step, within the same spectral range.
In practice, we have found this two-step 
procedure to be quite reliable, whereas 
simultaneous least-squares regression of
$c_{\eta}$ and $\varepsilon_{\infty}$ against 
the complex-valued
$\mathcal{E}^{\mathrm{intra}} (\omega)$ can yield
unphysical values for both parameters.

\subsection{Treatment of non-local  many-electron 
effects within the quasi-particle formalism:
Perturbative one-shot GW: G$_\mathbf{0}$W$_{\mathbf{0}}$}\label{sec:s2s1}

In order to calculate the aforementioned response functions, 
e.g. in Eq.~\ref{eq:s2e5}, a sufficiently complete set of
well-defined single-particle electronic states is required.
Density functional theory (DFT)~\cite{PhysRev.136.B864} is currently the almost-ubiquitously used approach for constructing   ground-state electronic structures of solids within its Kohn-Sham formalism (KS-DFT)~\cite{PhysRev.140.A1133}. 
However, the DFT is limited by the accuracy of available,
computationally feasible local and semi-local approximations for exchange and correlation~\cite{PhysRev.140.A1133,PhysRevB.21.5469,PhysRevB.33.8822,PhysRevB.46.6671}. 
Furthermore, the energy eigenvalues (band-structures) generated by
the Kohn-Sham mapping have no formal meaning in terms of 
electron addition or removal energies (except in certain 
well-documented instances), in spite of their being widely
interpreted as such.
The  RPA, although it is a true many-body approximation, 
is unable to build any electron or hole quasiparticle~\cite{landau1957theory,landau1957oscillations,landau1959theory} screening 
(e.g. electron-plasmon coupling) effects into an underlying 
Kohn-Sham eigensystem, as it treats only the screened
interaction between \emph{pairs} of such input particles.
The fact that the absence of non-local quantum many-body effects
in semi-local KS-DFT, 
and  absent explicit long-ranged exchange in particular, often leads to inaccurate descriptions of the electronic bands in solids
 is well reported for various material classes, such as  insulators and  semi-conductors~\cite{PhysRevLett.81.2312,PhysRevLett.91.256402,sottile2003response,PhysRevLett.88.066404,
PhysRevB.69.155112},  transition-metal oxides~\cite{PhysRevLett.74.2323,PhysRevB.55.13494,PhysRevB.60.15699}, and  metallic solids~\cite{godby1992unoccupied,0034-4885-61-3-002}.

In  noble metals,   the electronic bands that dominantly contribute  to low-energy spectra 
are  fully-filled $d^{10}$ bands tightly packed in a narrow energy window close to the Fermi level.
It has previously been found that these electronic bands are poorly described within  approximate KS-DFT for noble metals such as bulk Au~\cite{PhysRevB.86.125125}, and then that such  errors become 
more pronounced in  spectral simulations using the  RPA~\cite{0305-4608-14-1-013,doi:10.1021/cr200107z,PhysRevLett.88.016403,PhysRevB.66.115101,PhysRevB.66.161104}.
The formally correct approach to calculating band-structures from
 first principles is instead the quasiparticle (QP) formalism, 
fundamental to which is a mapping of the interacting many-body system  to a weakly-interacting  many-body system of virtual ones, namely the quasi-particles~\cite{pines1966elementary,aulbur2000quasiparticle}.
QP wave-functions and  corresponding energy levels can determined by self-consistently solving the QP equation
\begin{align}\label{eq:s2e15}
&\left[
-\frac{1}{2m_\mathrm{QP}}\nabla^2_i + v_{\mathrm{ext}}(\mathbf{r}) + v_{H}(\mathbf{r})\right.  \nonumber  \\
& +\left. \int\der{r'}\Sigma(\mathbf{r},\mathbf{r}',\omega)
\right]
\psi_i^\mathrm{QP}(\mathbf{r},\omega)
=\epsilon_i^\mathrm{QP}\psi_i^\mathrm{QP}(\mathbf{r},\omega),
\end{align}
where $\Sigma(\mathbf{r},\mathbf{r}',\omega)$ is the 
energy-dependent, non-local, and non-Hermitian self-energy. 
The resulting $\Psi_i^\textrm{QP}(\mathbf{r})$ and  $E_i^{QP}$ are the $i^{th}$ QP wave-function and the corresponding QP energy.
In  practice,  QP energies are more commonly 
calculated using nonetheless demanding
 many-body perturbation theory (MBPT)~\cite{fetter2012quantum} methods with Green's functions instead of explicit solution of Eq.~\eqref{eq:s2e15}.
%

%
%As seen in Eq.~\eqref{eq:s2e15}, the QP equation only differs from the KS equation  by the self-energy term instead of the locally defined xc-potential, where the KS equation can be recovered by defining the xc-potential as the local part of the self-energy at the static limit as
%%
%\begin{align}\label{eq:s2e16}
%v_{\mathrm{xc}}(\mathbf{r})=\lim_{\omega \rightarrow 0}\Sigma(\mathbf{r},\mathbf{r}',\omega)\delta(\mathbf{r}-\mathbf{r}').
%\end{align}
%

%\subsubsection*{}\label{sec:c5e3s3}

The GW approximation is the cornerstone of MBPT for electrons.
In GW, the self-energy is calculated in one iteration formally as 
\begin{align}\label{eq:s2e17}
\Sigma=\im G W,
\end{align}
where the product here is in real space and time.
The screened Coulomb interaction 
$W$ consistent with the GW approximation is that calculated 
 within the RPA for  a given Green's function G.
Self-consistent GW  is a computationally expensive approach that requires the solution of a  Dyson equation multiple times, and that involves the inversion of large, complex, and near-singular matrices.
Furthermore, it has been shown that it fails for systems with  over half-filled $3d$ bands~\cite{PhysRevB.47.15404}.
A more successful, further approximation is one-shot, non-self-consistent GW, or simply G$_0$W$_0$, which depends explicitly upon and input Green's electronic function $G_0$, and stops at the first iteration of the self-energy, 
$\Sigma=\im G_0 W_0$.%

In practice, this requires the choice of a suitable basis and,
when all that is of interest are the QP energies, it is
expedient to use the KS-DFT eigenbasis, on the grounds that  
the approximate KS-DFT density is usually reasonable, even if the KS-DFT eigenspectrum, represented in the form of G$_0$, is unphysical.
Assuming that $\langle \psi_i|\psi_i^\mathrm{QP}\rangle \approx 1$, 
the QP energies can furthermore be approximated as a first-order correction to the KS eigenvalues, as 
\begin{align}\label{eq:s2e20}
\epsilon_i^\mathrm{QP}&{}=\epsilon_i+
Z_i \langle\psi_i| \hat{\Sigma}\left(\epsilon_i\right) - \hat{v}_{\mathrm{xc}}|\psi_i \rangle,  \quad \mbox{where} \nonumber \\
Z_i&{}=\left[1- \left.\frac{\partial\Sigma'(\omega)}{\partial \omega}\right\vert_{\omega=\epsilon_i}\right]^{-1},
\end{align}
and $Z_i$ is called the QP re-normalization factor.
This factor can be thought of as the absolute value of the charge of the QP (e.g., of the electron and its screening cloud).
Here, $ \hat{v}_{\mathrm{xc}}$ is the approximate exchange-correlation
potential operator of KS-DFT.
The method described by this final step is called perturbative
G$_0$W$_0$, and it remains explicitly dependent on the  choice of approximate functional in KS-DFT.

In practice, in the KS wave-function basis, the non-interacting  single-particle Green function takes the form in the frequency domain,  for $t'>t$~\cite{marini2001optical},
\begin{align}\label{eq:s2e21}
G_0(\mathbf{r},\mathbf{r}',\omega)=&
2\im \sum_{i,\mathbf{k}}\psi_{i,\mathbf{k}}(\mathbf{r})\psi_{i,\mathbf{k}}(\mathbf{r}')  \nonumber \\
& \times \left(\frac{f_{i,\mathbf{k}}}{\omega-\epsilon_{i,\mathbf{k}}+\im\delta}+
\frac{1-f_{i,\mathbf{k}}}{\omega-\epsilon_{i,\mathbf{k}}-\im\delta} \right)
\end{align}
and, once $G_0$ is obtained,  $W_0$ is calculated by using Dyson's equation starting from the bare Coulomb interaction with
\begin{align}\label{eq:s2e22}
W_0=\varepsilon^{-1}v,
\end{align}
where $W^0$ is a function of $\omega$ through the inverse  dielectric function $\varepsilon^{-1}$.
In  reciprocal space, $W_0$ is expressed by Eq.~\eqref{eq:s2e4} and Eq.~\eqref{eq:s2e8} as the direct product
\begin{align}\label{eq:s2e23}
W_{0 \mathbf{G},\mathbf{G}}(\mathbf{q},\omega)=\varepsilon^{-1}_{\mathbf{G},\mathbf{G}'}(\mathbf{q},\omega)\frac{4\pi}{|\mathbf{q}+\mathbf{G}||\mathbf{q}+\mathbf{G}'|}.
\end{align}
Computationally, the inversion of the dielectric function, which is a large matrix with  frequency-dependent complex entities, is troublesome.  
Hence, the frequency-dependent complex entities are approximated by  Lorentzian peaks within the  plasmon-pole approximation (PPA)~\cite{PhysRevB.34.5390,PhysRevLett.62.1169}.
The idea behind the PPA is to replace the single-particle transitions making up $\varepsilon^{-1}_\mathbf{G,G'}(\mathbf{q},\omega)$,  which increase in number with the square of the system size, with a smaller number of effective plasmon modes.
Thus $\varepsilon^{-1}$ is approximated in practice  via
\begin{align}\label{eq:s2e24}
\varepsilon^{-1}_{\mathbf{G},\mathbf{G}'}(\mathbf{q},\omega)=\delta_{\mathbf{G},\mathbf{G}'}
+\sum_\lambda \frac{\Omega^\lambda_{\mathbf{G},\mathbf{G}'}}{\omega^2-\left(\bar{\omega}^\lambda_{\mathbf{G},\mathbf{G}'}\right)^2},
\end{align}
where $\Omega_{\mathbf{G},\mathbf{G}'}$ and $\bar{\omega}^\lambda_{\mathbf{G},\mathbf{G}'}$ are the strength and the frequency of  plasmons fitted to the RPA inverse dielectric function.
A second advantage of the PPA is that 
analytical formulae for $\Sigma$ are available in the frequency
domain.

\section{Computational details}

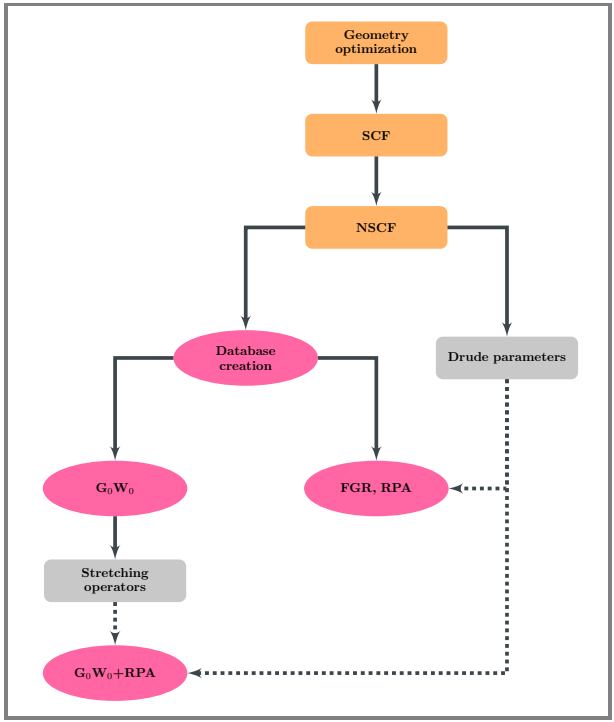
\begin{figure}[H]
\centering
\tcbox[sharp corners, boxsep=0.0mm, boxrule=0.5mm,  colframe=gray, colback=white]{
\resizebox{0.4\textwidth}{!}{
\begin{tikzpicture}
    % Place nodes
    \node [block] (gop) {\textbf{Geometry\\ optimization}};
    \node [block, below of=gop,node distance=2.5cm] (scf) {\textbf{SCF}};
    \node [block, below of=scf,node distance=2.5cm] (nscf) {\textbf{NSCF}};
    \node [decision, below right of=nscf,node distance=5cm] (dp) {\textbf{Drude parameters}};
    \node [cloud, below left of=nscf,node distance=5cm] (data) {\textbf{Database creation}};
    
	 \node [cloud, below right of=data, node distance=5cm] (fgr) {\textbf{FGR, RPA}};    
    
    \node [cloud, below  left of=data,node distance=5cm] (gw) {\textbf{G$_0$W$_0$}};
    \node [decision, below of=gw,node distance=2.5cm] (sc) {\textbf{Stretching operators}};
    \node [cloud, below of=sc, node distance=2.5cm] (gwrpa) {\textbf{G$_0$W$_0$+RPA}};
    
    % Draw edges
    
    \path [line] (gop) -- (scf);
    \path [line] (scf) -- (nscf);
    \path [line] (nscf) -| (data);
    \path [line] (nscf) -| (dp);
    \path [line] (data) -|(gw);
    \path [line] (gw) -- (sc);
    \path [line,dashed] (dp) |- (gwrpa);
     \path [line,dashed] (dp) |- (fgr);
    \path [line] (data) -|(fgr);
    \path [line,dashed] (sc) -- (gwrpa);
\end{tikzpicture}}}
\caption{The work-flow followed here to calculate the optical spectra of Au\td{x}Ag\td{y}Cu\td{1-x-y} using Quantum ESPRESSO (QE) and Yambo. Color codes are assigned to each software with QE as orange, Yambo as pink, and in-house  code as gray.}
\label{fig:c6-wf}
\end{figure}
Geometry optimization, self-consistent field (SCF) and non-self-consistent field (NSCF) simulations were performed  using the  Quantum ESPRESSO software (QE)~\cite{0953-8984-21-39-395502,0953-8984-29-46-465901}. 
For this, norm-conserving  PBE pseudo-potentials 
were produced using the pseudo-potential generator OPIUM~\cite{opium}. 
The initial crystallographic information for bulk Au,  Ag, and Cu  in their FCC structures were adopted from X-ray diffraction data at  1072 K from Ref.~\citenum{Suh1988}, for consistency with choice of smearing parameter for the Marzari-Vanderbilt cold smearing~\cite{PhysRevLett.82.3296}, namely $0.1$~eV.
Full geometry relaxations were then performed with variable cell parameters at over-converged plane-wave cutoff energies (E\td{cut} $= 75$~Ha) with 
automatically generated
$4 \times 4 \times 4 $  uniform Monkhorst-Pack Brillouin zone sampling, without imposing any crystal symmetry. 

For band-structure calculations, we moved down to a plane-wave cutoff energy of E\td{cut}~$= 25$~Ha, 
which is sufficient to attain a total-energy tolerance of
$\Delta E_\mathrm{tot} \le 0.001$ Ha per atom, but up to  a uniform
 Brillouin zone sampling of  $16 \times 16 \times 16$ at the NSCF level, 
which was necessary to converge the expectation value of the exchange
 self-energy.
Experimental spectra have large inter-band smearing 
$\Gamma$ values due to finite temperature effects and  impurities in 
practice~\cite{gurzhi1958theory,gurzhi1959mutual,fox2002optical}. 
Hence, a Lorentzian smearing parameter of full-width $0.2$ eV was adopted for spectral simulation with the Yambo code~\cite{MARINI20091392}. 
Final simulations with a common set of parameters were performed on Au, Ag, and Cu for benchmarking the various levels of theory studied, against experimental spectra.
The work-flow for simulations of spectra with FGR, RPA and G\td{0}W\td{0}+RPA is illustrated in Fig.~\ref{fig:c6-wf}.

\section{Work-flow optimization and 
benchmarking on pure metals: Au, Ag, and Cu}

For noble metals, we have found that even a converged inaccurate semi-local KS-DFT description of the relevant quasiparticle bands requires demanding run-time parameters and accurate pseudopotentials.
Therefore, when larger crystallographic unit cells are 
of interest such as for the alloys central to this work,
even perturbative G$_0$W$_0$ becomes computationally impractical, 
and we cannot routinely 
go very far beyond KS-DFT in terms of computational overhead.
Hence, here, we have pursued an intermediate, 
compromise approach in which non-local quasiparticle screening effects are incorporated approximately, in a scaleable manner which
incurs a minimum additional computational cost that is insignificant compared to 
the final  RPA calculation for the spectrum.

In principle, \gwr spectra  require a full QP band-structure as a starting point. 
To to obtain the QP band-structure, we would need to evaluate  QP energies for each point in the Brillouin zone. for every band.
For a given band and point in the Brillouin zone $\{i,\mathbf{k}\}$, such an operation consists of  summing throughout  the Brillouin zone  and over bands to determine G\td{0}, as well as  W\td{0} through the inverse dynamic dielectric function for the self-energy operator in Eq.~\eqref{eq:s2e20}.
Such a task is highly demanding both in terms of CPU hours as well as RAM. 
This operation needs to load all information about KS wave-functions in each processor unit, when using Yambo.
Since   our eventual goal here is to construct optical  spectra rather than QP bands,  an averaged stretching to underlying KS band-structure via stretching operators close to the Fermi level is sufficient, as well as more feasible for the spectral range of interest.
The idea of stretching operators is to approximate QP energies as  linear-functions of the  KS  energies, and for metals in particular we have simply
\begin{align}\label{eq:s4e25}
\epsilon_\mathrm{v}^\mathrm{QP}=s_\mathrm{v} \epsilon_\mathrm{v},\hspace{0.5cm}\mathrm{and}\hspace{0.5cm}
\epsilon_\mathrm{c}^\mathrm{QP}=s_\mathrm{c} \epsilon_\mathrm{c},
\end{align}
where $s_\mathrm{v}$ and $s_\mathrm{c}$ are separate stretching factors for the  valence and conduction bands, respectively. 
Such an approach introduces  averaged corrections to the valence and conduction bands around  Fermi level for the missing non-local electronic exchange-correlation
effects, whilst keeping the Fermi level fixed. 
\begin{figure}[t]
\centering
\subfloat[Au  \label{fig:au-sciss}]{
\includegraphics[width=0.22\textwidth]{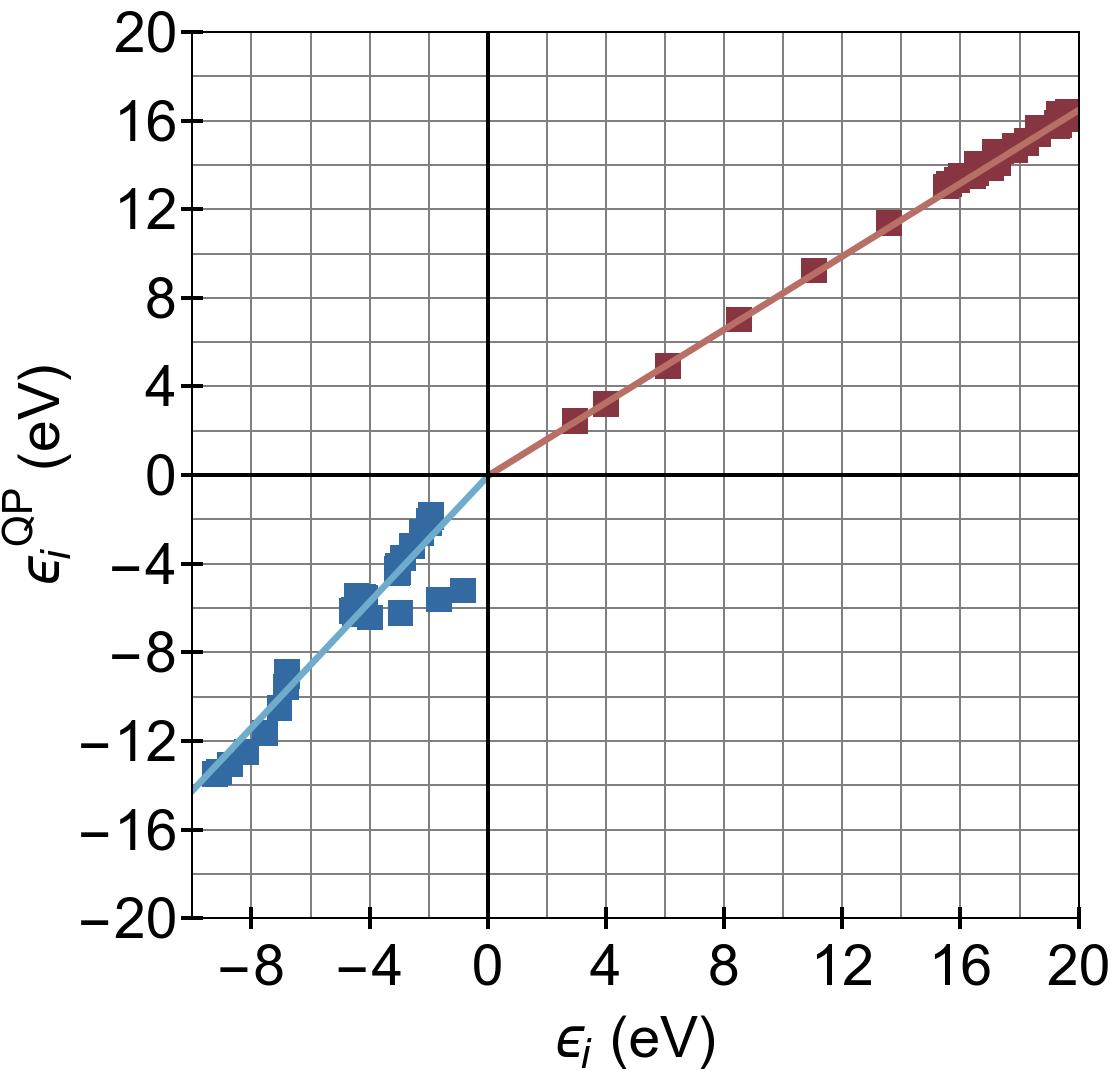}
}\hfill
\subfloat[Ag  \label{fig:ag-sciss}]{
\includegraphics[width=0.22\textwidth]{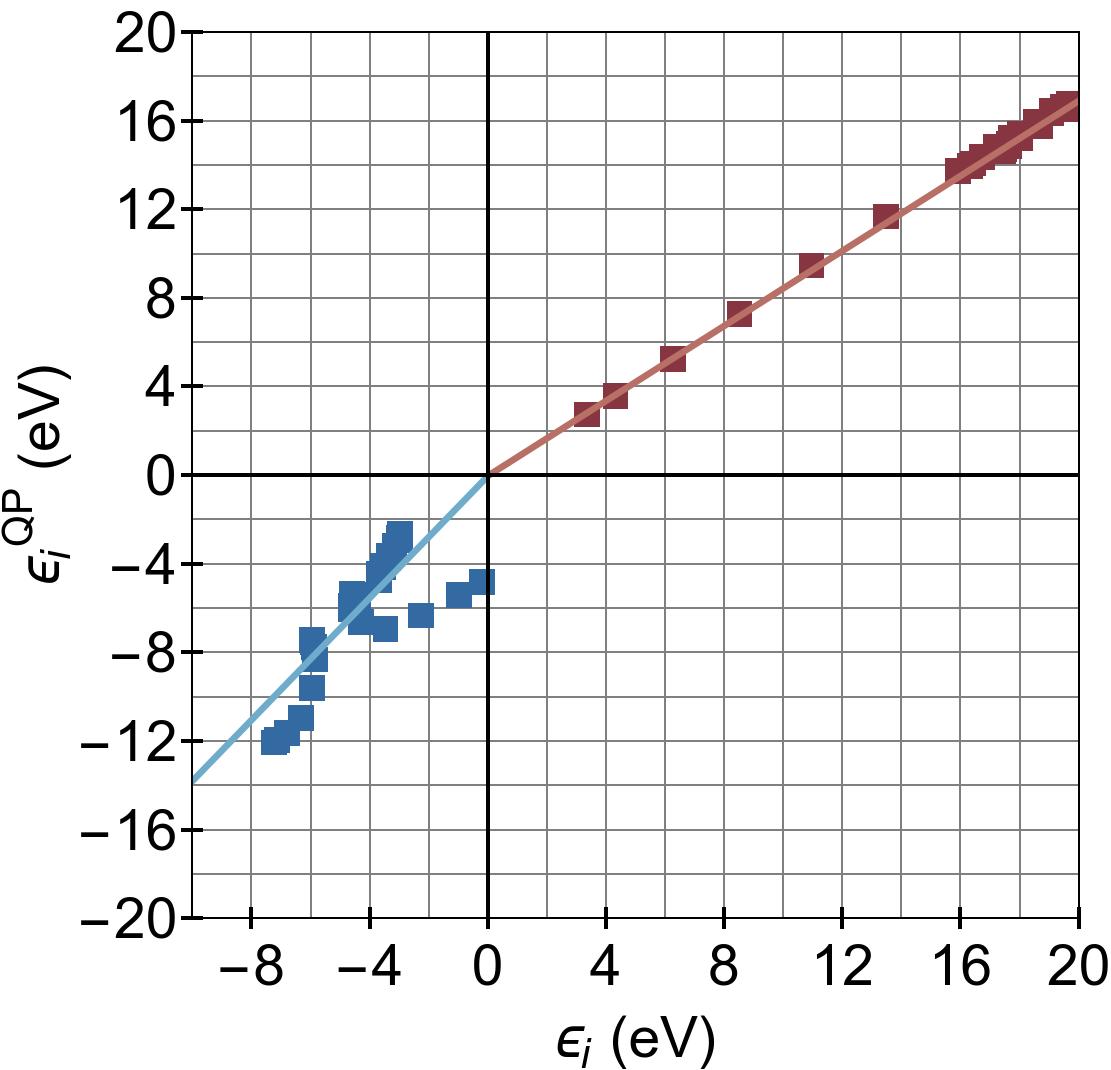}
}\hfill
\subfloat[Cu  \label{fig:cu-sciss}]{
\includegraphics[width=0.22\textwidth]{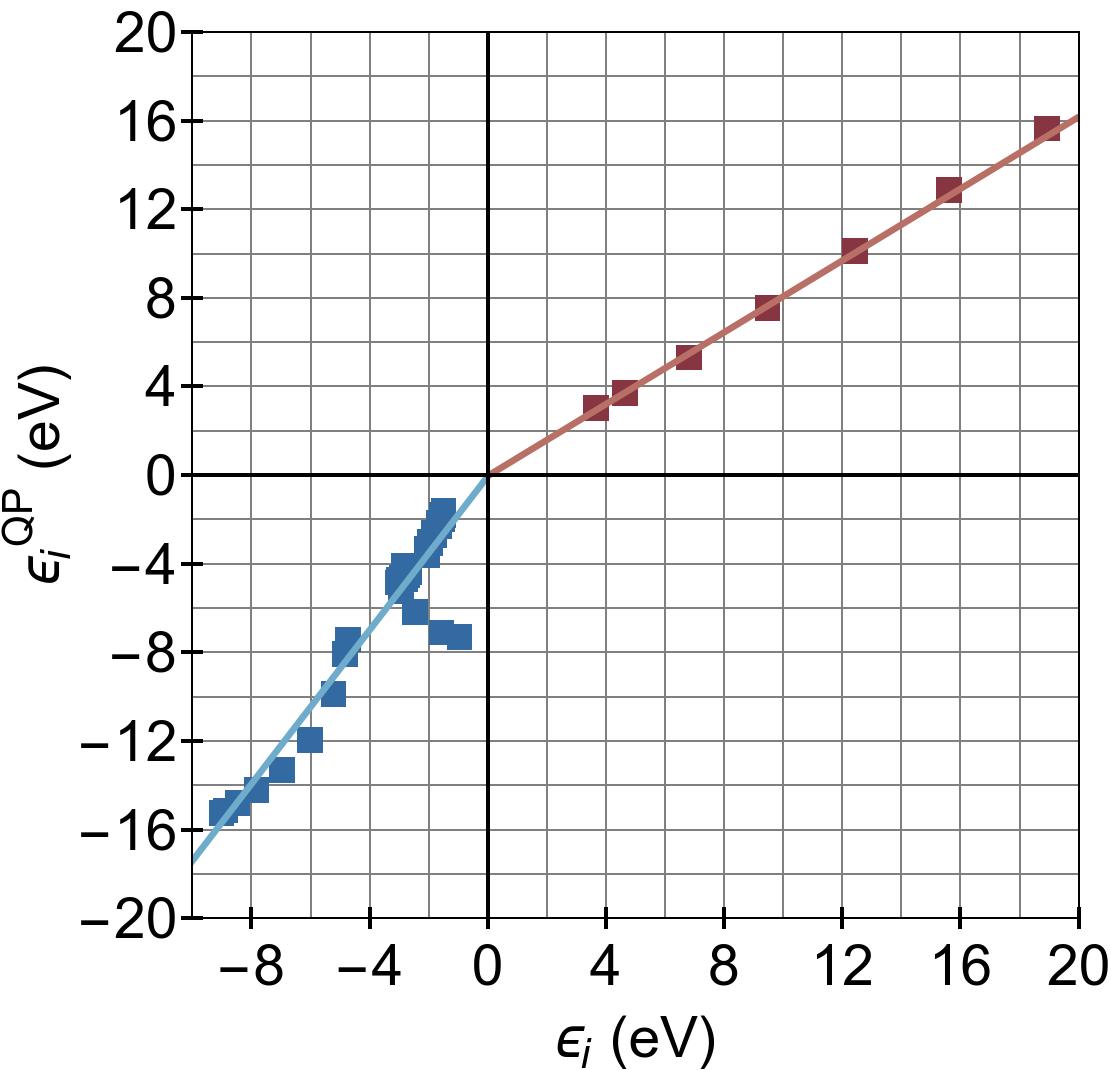}
}
\caption{Determination of the stretching operators for Au, Ag, and Cu by means linear fitting to KS-DFT vs. QP eigen-energies using 10 points around the $\Gamma$ point for 6 valence (blue points) and 6 conduction (red points) bands. The Fermi levels are shifted to $0$~eV in each case.}
\label{fig:nob-sciss}
\end{figure}
The stretching operators were determined by linear regression on $\epsilon_i$ vs. $\epsilon_i^\textrm{QP}$.
For  pure metals, QP energies were calculated for 6 valence bands and 6 conduction bands at 10 points at and around $\Gamma$,  and the stretching operators were determined by linear fitting as shown in Fig.~\ref{fig:nob-sciss} with the values listed in Table~\ref{tab:c6t3}.
In Fig.~\ref{fig:nob-sciss}, two  branches are observed in the valence manifolds. 
These distinctive branches are due to different  non-local exchange contributions to $s$ and $d$-bands, but nonetheless a single stretching factor proved to be 
adequate.

\begin{table}[H]
\renewcommand{\arraystretch}{1.6} \setlength{\tabcolsep}{15pt}
\begin{center}
{\small
\begin{tabular}{lcc}
\hline \hline 
 & $s_\mathrm{v}$ & $s_\mathrm{c}$  \\ \hline
 Au & 1.419797  &  0.825253  \\
Ag & 1.376302  &  0.846172  \\
Cu &  1.735804  & 0.809883 \\
 \hline \hline
\end{tabular}}
\end{center}
\caption{Calculated values for  stretching operators used to stretch KS band-structures in order to imitate perturbative
G$_0$W$_0$ QP band-structures for  pure Au, Ag, and Cu.}
\label{tab:c6t3}
\end{table}

\begin{figure}[t]
\centering

\subfloat[Au \label{fig:au-band}]{\includegraphics[width=0.3\textwidth]{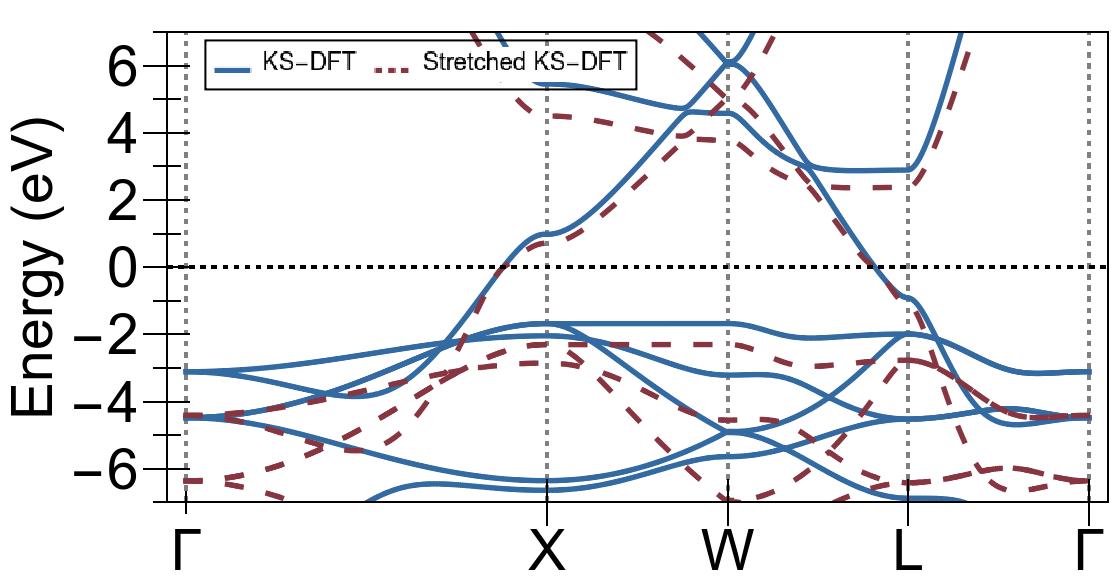}}

\subfloat[Ag \label{fig:ag-band}]{\includegraphics[width=0.3\textwidth]{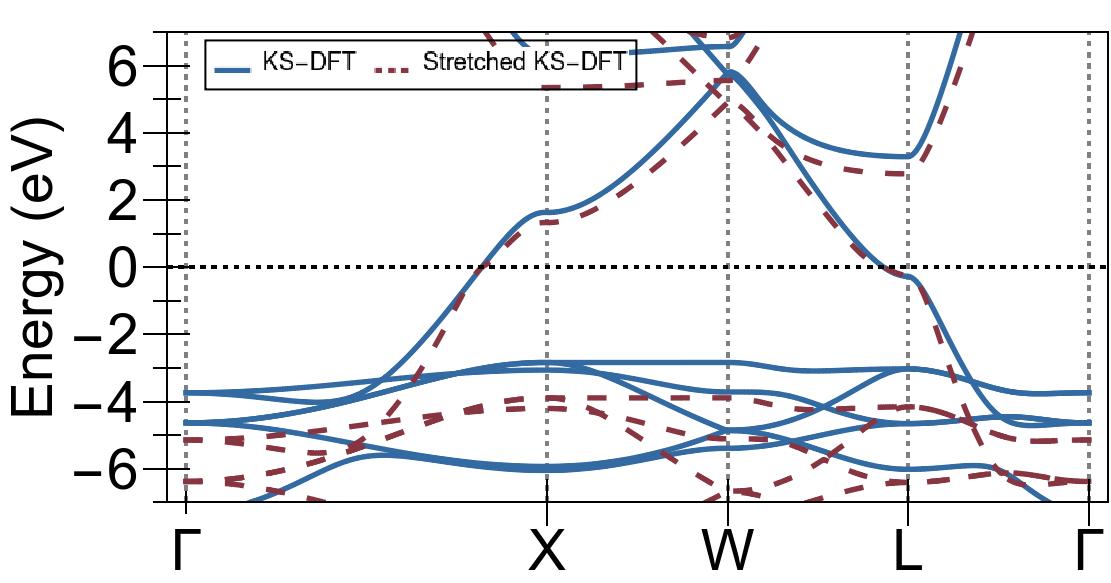}}

\subfloat[Cu \label{fig:cu-band}]{\includegraphics[width=0.3\textwidth]{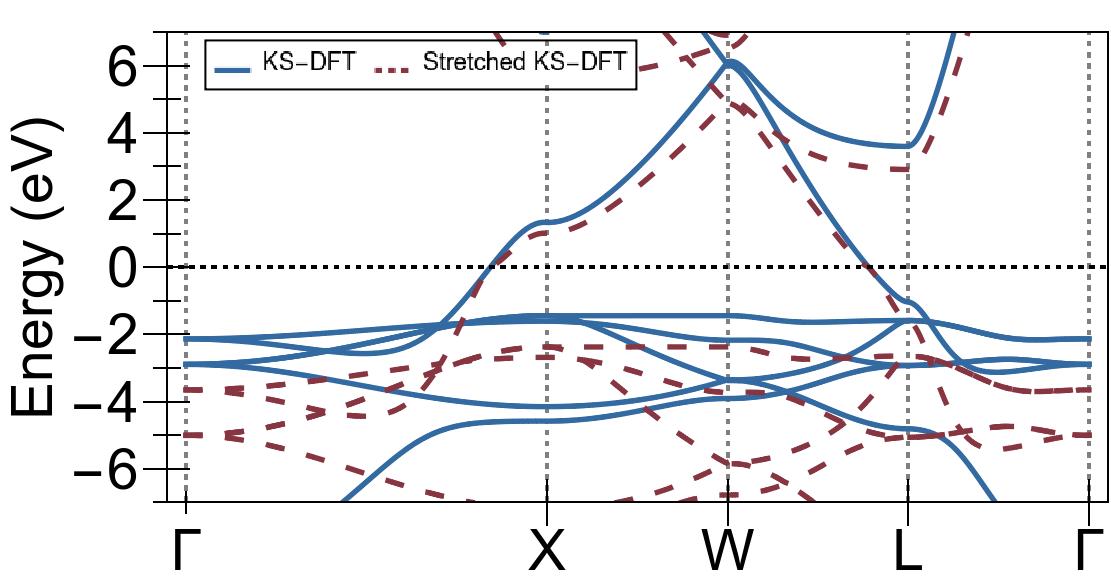}}

\caption{KS and QP band-structures for Au, Ag, and Cu along a  high-symmetry path with a 0.001~a$_0^{-1}$ step size.}
\label{fig:nob-band}
\end{figure}
The stretching operators modify the KS band-structure as  shown in Fig.~\ref{fig:nob-band}.
The Fermi-Dirac distribution for the chosen electronic temperature was used to interpolate between the distinct stretching parameters for the valence and conduction bands. 
KS-DFT tends to excessively flatten the fully filled $d$-bands due to an absence of attractive  non-local exchange.
In a sense,  the stretching operator approximately corrects the  dispersions of the bands, particularly for the occupied $d$-bands but also for the half-occupied $s$-band, which is made less dispersive.
In Fig.~\ref{fig:cu-band} for Cu,  the bands close to the Fermi level are more narrowly packed and flattened compared to those of Ag and Au cases, reflecting the fact that  the error in the KS-DFT treatment, and hence $s_v^{Cu}$, is the largest amongst the three.
The inverse behaviour is seen, however, for the conduction stretching operators of pure metals, which is due to correlation effects only,  as Ag has the largest stretching factor  in the conduction manifold, albeit that the differences between the metals
are less pronounced here.

\subsection{Drude parameters for pure metals}
Before constructing optical spectra, the Drude parameters are needed for the intra-band part of the dielectric function in Eq.~\eqref{eq:s2e14}.
Our first step is to calculate the Drude plasmon energies using Eq.~\eqref{eq:s2e12}.
For this purpose, the energies of bands crossing the Fermi level at each k-point were extracted from the output of NSCF calculations and interpolated on a fine grid in the Brillouin zone (601 points in each reciprocal-space direction),  and the Fermi surface was located on this grid with a $\pm 0.01$ eV tolerance for each system.
Then,  the square of the Fermi velocities, averaged over the Fermi surface, were calculated by means of Eq.~\eqref{eq:s2e11}.
%
%The average Fermi velocity itself is given by
%%
%\begin{gather}\label{eq:s4e26}
%v_\mathrm{F}^\mathrm{KS}  = \sqrt{\langle v^2(E_\mathrm{F})\rangle }=\left[\frac{\sum_i\int_{S_\mathrm{F}}\der{S} v^2_i(\mathbf{k}) }{\int_{S_\mathrm{F}}\der{S}}\right]^{1/2}, \nonumber \\
%%
%v^2_i(\mathbf{k})=\left|\frac{\partial \epsilon(\mathbf{k})}{\partial \mathbf{k}}\right|^2, 
%\end{gather}
%%
%where $S_F$ is the Fermi surface.
%
The calculated Fermi velocities for pure Au, Ag, and Cu are listed in Table~\ref{tab:c6t4} along with results of Ref.~\citenum{doi:10.1063/1.4942216}, which uses a similar procedure with an extremely dense Brillouin zone sampling as $200 \times 200 \times 200$ rather than interpolating,  as we have.
Furthermore, the average Fermi velocity magnitudes for QP band-structures were approximated by applying the geometric averages 
(more appropriate to simulate intra-band response than the arithmetic mean) 
of the  valence and conduction stretching operators, specifically
using the formula
\begin{align}\label{eq:s4e27}
v^\mathrm{G_0W_0}
(E_\mathrm{F})=\sqrt{s_\mathrm{v}s_\mathrm{c}}v^\mathrm{KS}
(E_\mathrm{F}).
\end{align}

\begin{table}[H]
\renewcommand{\arraystretch}{1.8} \setlength{\tabcolsep}{12pt}
\begin{center}
{\small
\begin{tabular}{lccc}
\hline \hline 
 & $v^\mathrm{KS} (E_\mathrm{F})$ & $v^\mathrm{G_0W_0} (E_\mathrm{F})$  & $v^\mathrm{Gall} (E_\mathrm{F})$~\cite{ doi:10.1063/1.4942216} \\ \hline
 Au & 13.476  & 14.588  &13.82  \\
Ag & 13.912  & 15.014  &14.48 \\
Cu &  10.794 & 12.798  &11.09\\
 \hline \hline
\end{tabular}}
\end{center}
\caption[The average Fermi velocities for pure Au, Ag, and Cu.]{The average Fermi velocities (in $\times 10^5$ m/s units) for pure Au, Ag, and Cu. We have included results from  the work of Gall~\cite{doi:10.1063/1.4942216}(the third column), who evaluated the Fermi velocities using  a $200 \times 200 \times 200 $ Brillouin zone sampling.}
\label{tab:c6t4}
\end{table}
Our averaged KS-DFT Fermi velocities  slightly underestimate those of
Ref.~\onlinecite{doi:10.1063/1.4942216}, as shown in the third column of Table~\ref{tab:c6t4}, and the origin of this discrepancy is not evident.
Considering  the much smaller computational cost of our interpolation scheme, our values are very reasonable estimates of the Drude plasmon energies.
Lastly, the DOS based on the KS band-structure and QP band-structure were extracted by applying a $0.1$ eV   Lorentzian broadening  using a post-processing tool of Yambo.
Using Eq.~\eqref{eq:s2e12},  the Drude plasmon energies were estimated for  the KS band-structures and  QP band-structures with their respective DOS at the Fermi level.

\begin{figure}[h!]
\centering
\tcbox[sharp corners, boxsep=0.0mm, boxrule=0.5mm,  colframe=gray, colback=white]{
\resizebox{0.4\textwidth}{!}{\begin{tikzpicture}
%\draw [help lines] (0,0) grid (7,5);

\node [rectangle, draw=sangria, fill=sangria!10,line width=0.5mm, minimum height=1cm, text width=4cm,text centered, rounded corners] (ksband) {KS band-structure};

\node [rectangle, draw=sangria, fill=sangria!10,line width=0.5mm, minimum height=1cm, text width=1.5cm,text centered, rounded corners,below right of=ksband,node distance=3cm](vf) {$\langle v^2(E_\mathrm{F}) \rangle$};

\node [rectangle, draw=sangria, fill=sangria!10, line width=0.5mm, minimum height=1cm, text width=1.2cm,text centered, rounded corners,above right of=ksband,node distance=3cm](dos) {$N(E_\mathrm{F})$};

\node [rectangle, draw=sangria, fill=sangria!10,line width=0.5mm, minimum height=1cm, text width=2cm,text centered, rounded corners, right of=ksband,node distance=4cm](wp) {$\omega_\mathrm{p}, \; \omega_\mathrm{p}^\mathrm{G_0W_0}$};

\node [rectangle, draw=prussianblue, fill=prussianblue!10,line width=0.5mm, minimum height=1cm, text width=2.5cm,text centered, rounded corners, right of=wp,node distance=3cm](ep2) {$\ep_2^{\mathrm{intra}}(\omega, \omega_\mathrm{p},\eta_\mathrm{p})$};

\node [rectangle, draw=prussianblue,fill=prussianblue!10, line width=0.5mm, minimum height=1cm, text width=1cm,text centered, rounded corners, right of=ep2,node distance=2.5cm](eta) {$\eta_\mathrm{p}$};

\node [rectangle, draw=phthalogreen,fill=phthalogreen!10, line width=0.5mm, minimum height=1cm, text width=3.75cm,text centered, rounded corners, below of=ep2,node distance=2cm](ep1) {$\ep_1^{\mathrm{intra}}(\omega, \omega_\mathrm{p},\eta_\mathrm{p},\varepsilon_\infty)$};

\node [rectangle, draw=phthalogreen,fill=phthalogreen!10, line width=0.5mm, minimum height=1cm, text width=1cm,text centered, rounded corners, below of=ep1,node distance=2cm](epinf) {$\varepsilon_\infty$};

\draw [line, line width=0.5mm] (ksband) -- (vf) ;
\draw [line, line width=0.5mm] (ksband) -- (dos) ;

\draw [line,line width=0.5mm] (dos) -- (wp) node [midway, above, sloped] (TextNode) {$s_\mathrm{v},s_\mathrm{c}$};
\draw [line,line width=0.5mm] (vf) -- (wp) node [midway, below, sloped] (TextNode) {$s_\mathrm{v},s_\mathrm{c}$};

\draw [line, line width=0.5mm] (wp) -- (ep2) ;
\draw [line, line width=0.5mm] (ep2) -- (eta) ;

\draw [line, line width=0.5mm] (wp) -- (ep1) ;
\draw [line, line width=0.5mm] (eta) -- (ep1) ;

\draw [line, line width=0.5mm] (ep1) -- (epinf) ;

\end{tikzpicture}}}
\caption{A schematic of the semi-empirical method used to obtain  Drude parameters for pure metals.}
\label{fig:c6-semiemp}
\end{figure}
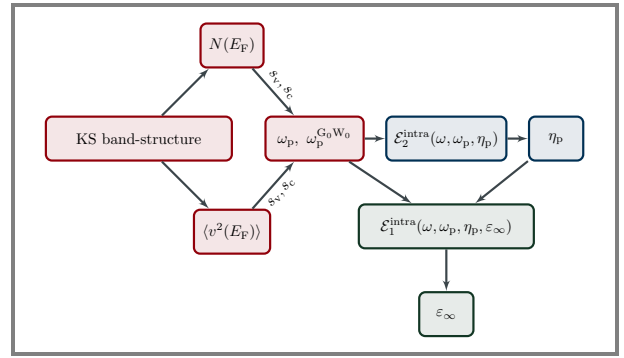

The next step is to approximate the inverse life-time of the  Drude plasmon, as well as the electric permittivity in the infinite-frequency limit, using our semi-empirical approach  illustrated in Fig.~\ref{fig:c6-semiemp}.
Scaling factors for the inverse life-times $\eta_\mathrm{p}$ in Eq.~\ref{eq:s2e13} were determined by fitting the imaginary part of the dielectric function to the experimental curves in Ref.~\citenum{Babar:15}, and then $\varepsilon_\infty$ values were determined by fitting the real part of the dielectric function with sets of $\{\omega_\mathrm{p},\eta_\mathrm{p}\}$ to the same experimental curves for Au, Ag, and Cu.
The resulting values for FGR, RPA and \gwr are summarized in the Supporting Information (SI).

\begin{figure}[H]
\centering
\includegraphics[width=0.23\textwidth]{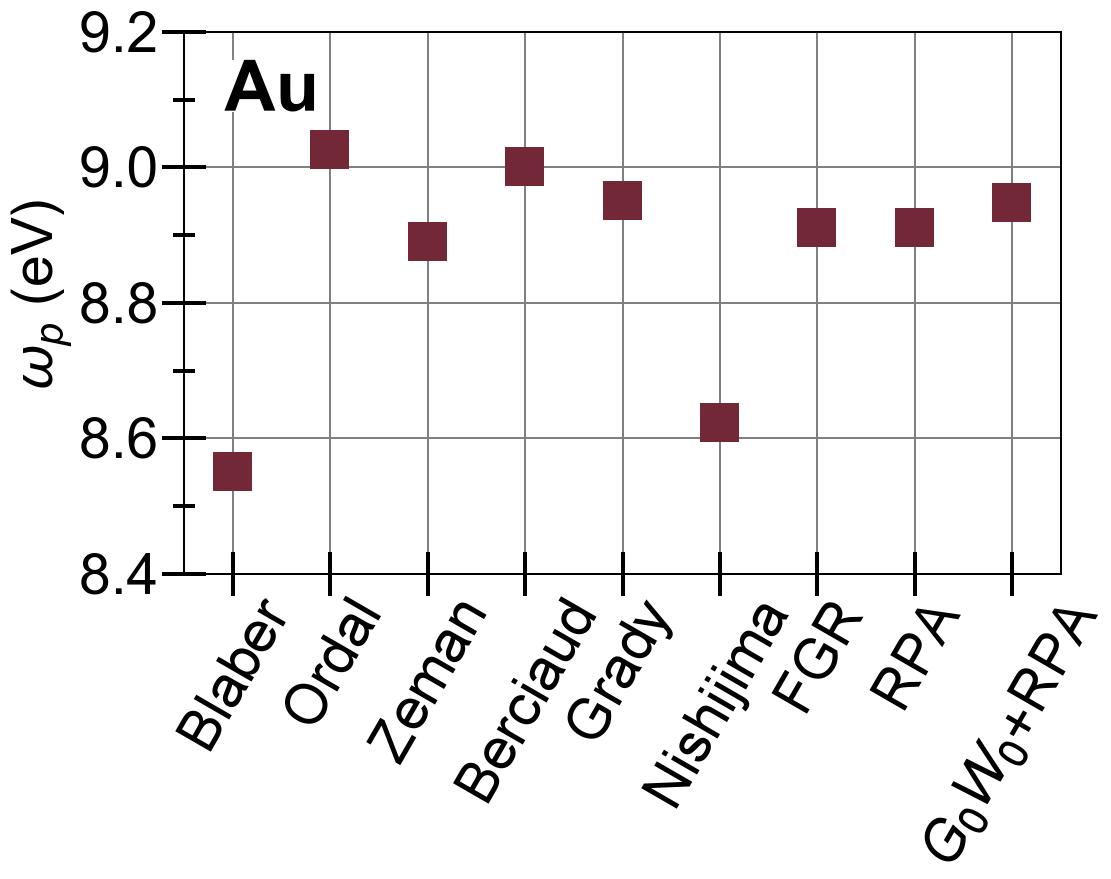}
\includegraphics[width=0.23\textwidth]{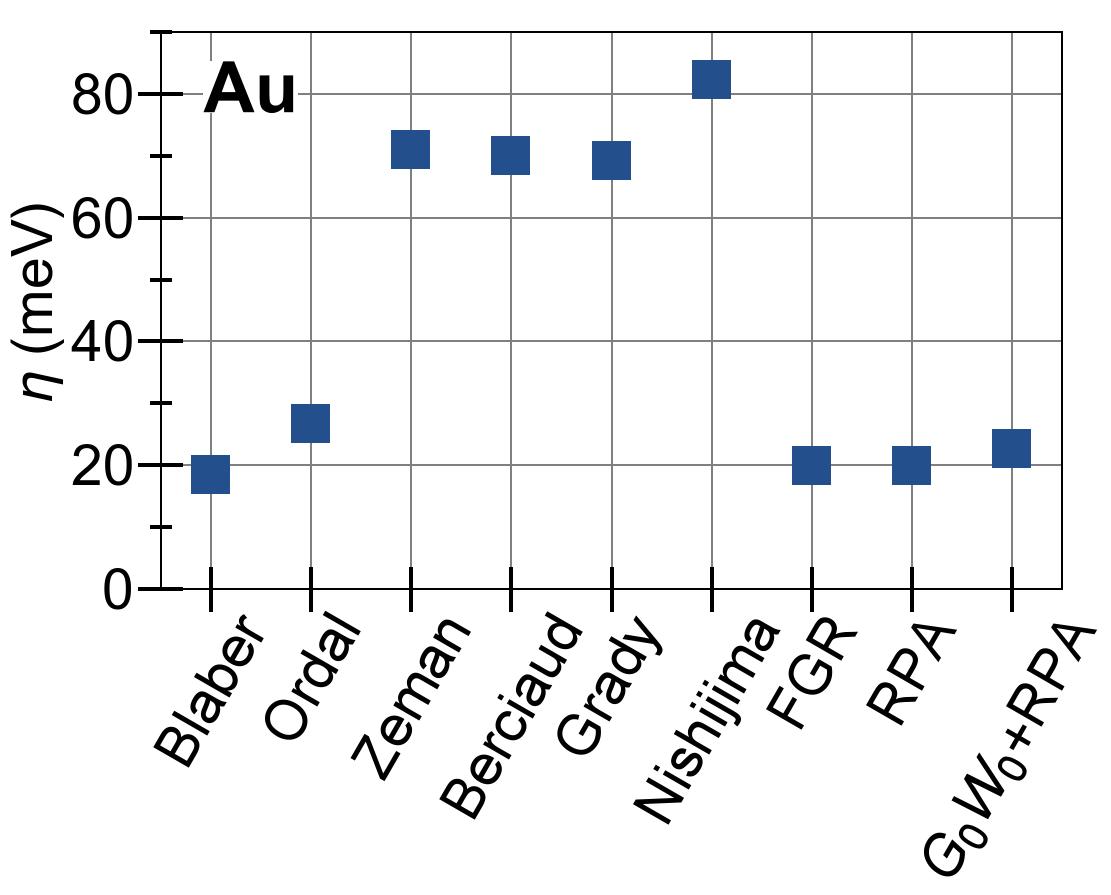}  \vspace{0.25cm}

\includegraphics[width=0.23\textwidth]{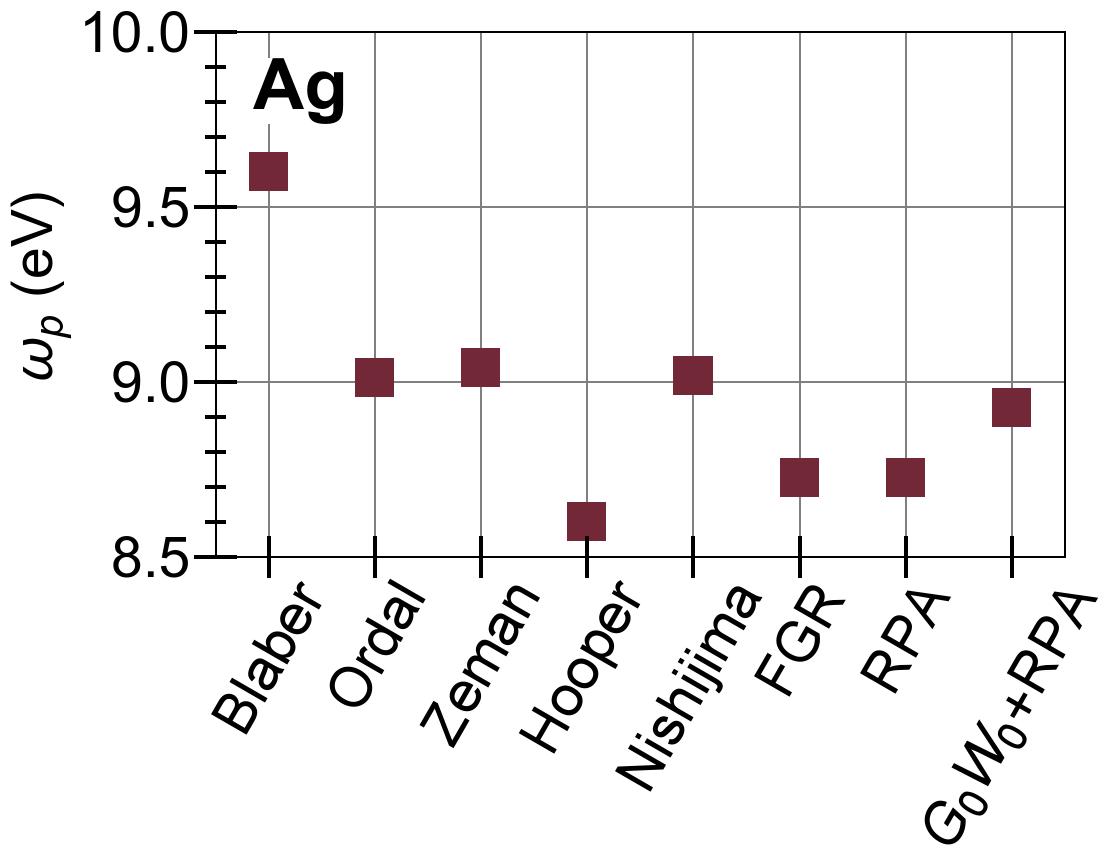}
\includegraphics[width=0.23\textwidth]{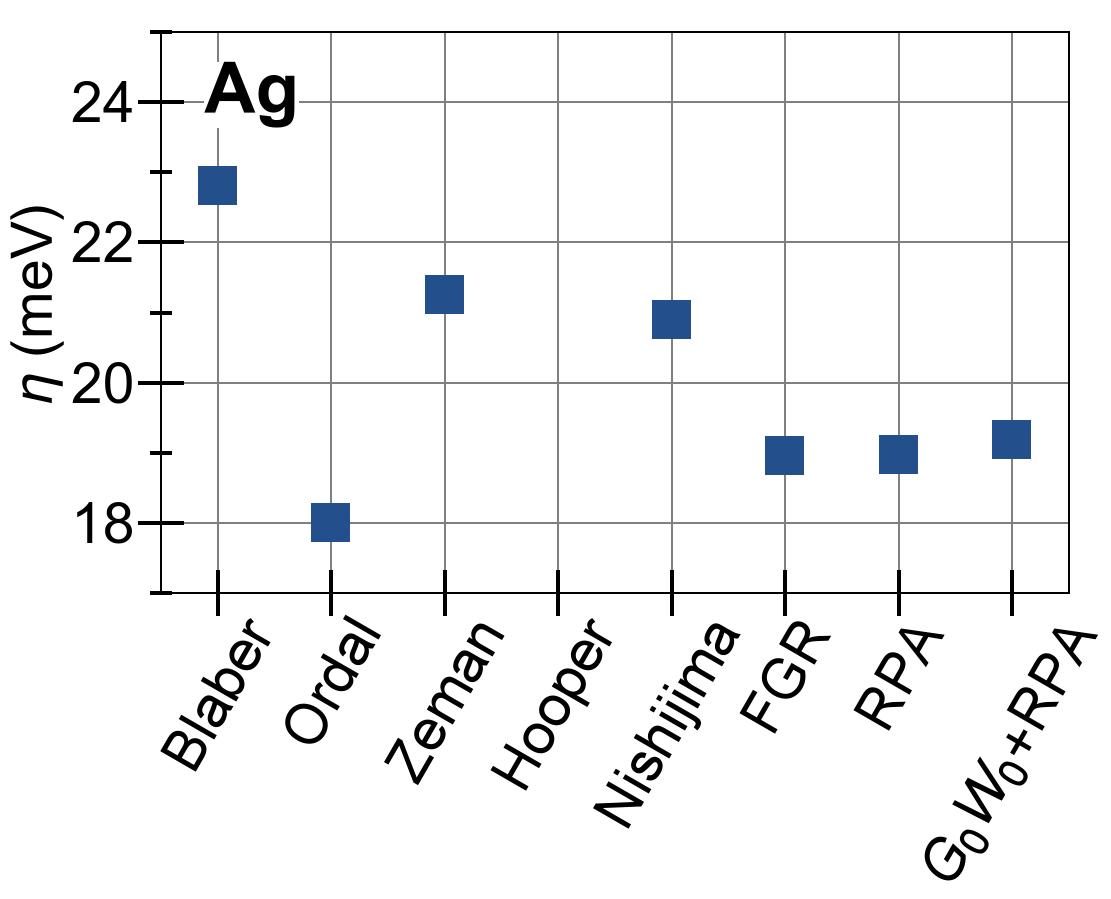} \vspace{0.25cm}

\includegraphics[width=0.23\textwidth]{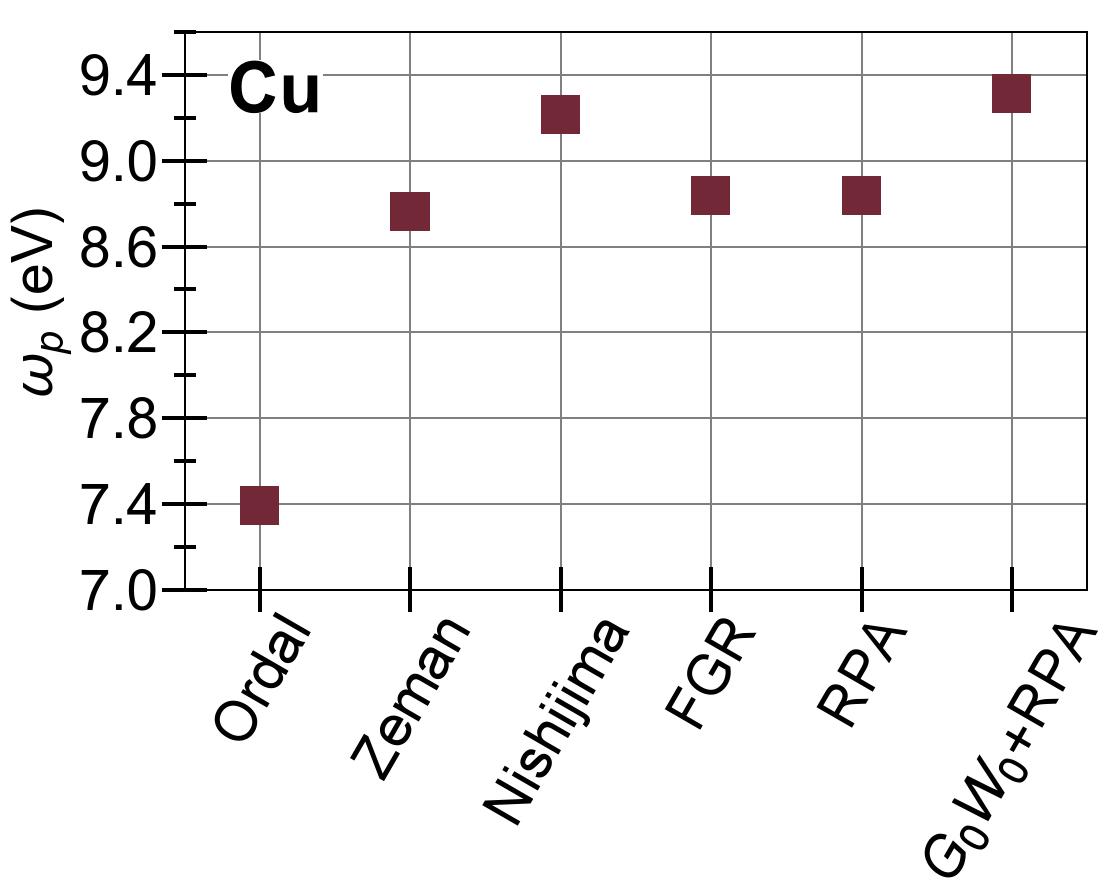}
\includegraphics[width=0.23\textwidth]{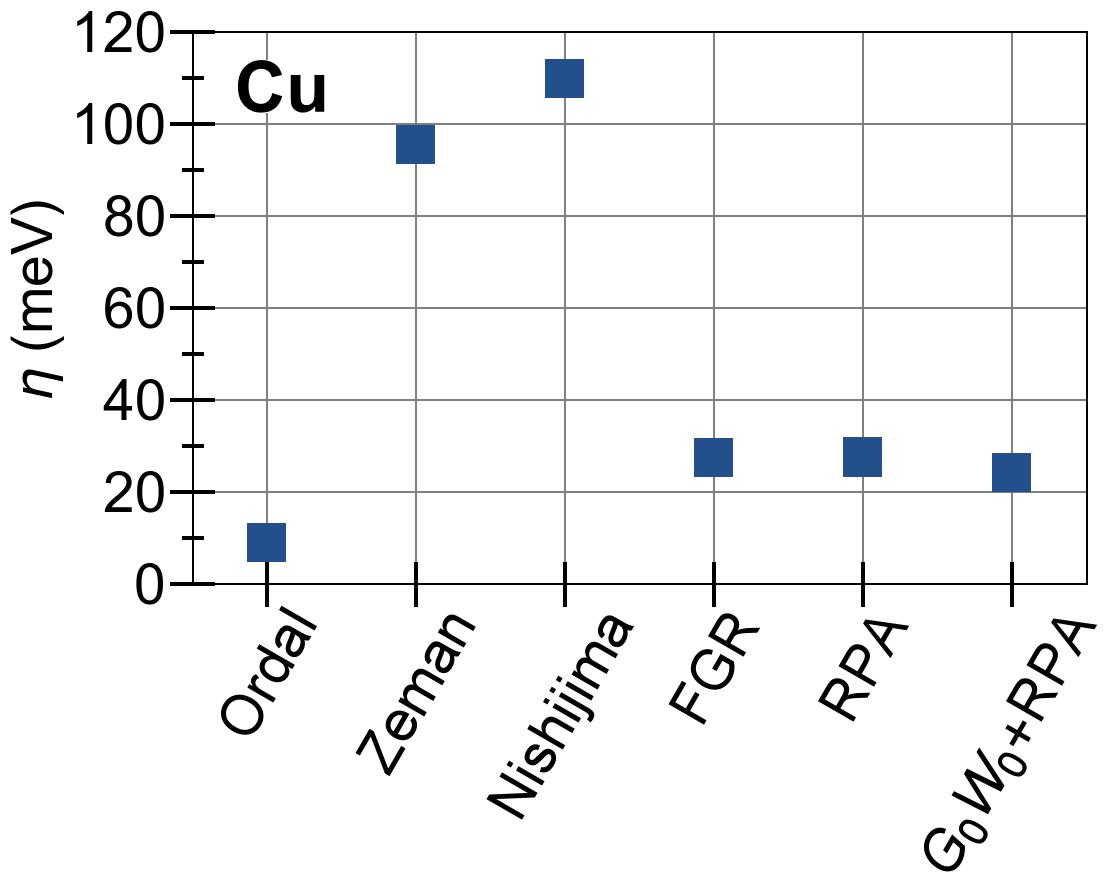}

\caption[Comparison of computed Drude parameters against experiment for Au, Ag, and Cu.]{ Comparison of the Drude plasmon energies and inverse life-times against commonly-used experimental values for these parameters, specifically from Blaber~\cite{doi:10.1021/jp810808h}, Ordal~\cite{Ordal:85}, Zeman~\cite{doi:10.1021/j100287a028}, Berciaud~\cite{doi:10.1021/nl050062t}, Grady~\cite{Grady2004167}, Hooper~\cite{PhysRevB.65.165432}, and Nishijima~\cite{hashimoto2016ag}.}
\label{fig:nob-dp-comp}
\end{figure}
In experimental studies, the Drude parameters are commonly determined by least-squares fitting  of the  Drude-Lorentz model in Eq.~\eqref{eq:s2e14} to  measurements in the IR spectral range.
Such measurements are highly sensitive to  experimental details, and
the resulting literature for the Drude parameters  is not in good consensus as shown  Fig.~\ref{fig:nob-dp-comp}.
This figure shows that, despite  the relative simplicity of the approaches  adopted in this work, our Drude parameters are comparable with experimental predictions
(particularly those of Ordal~\cite{Ordal:85}). 

\subsection{Optical spectra of pure metals}
\begin{figure}[t]
\centering
\includegraphics[width=0.25\textwidth]{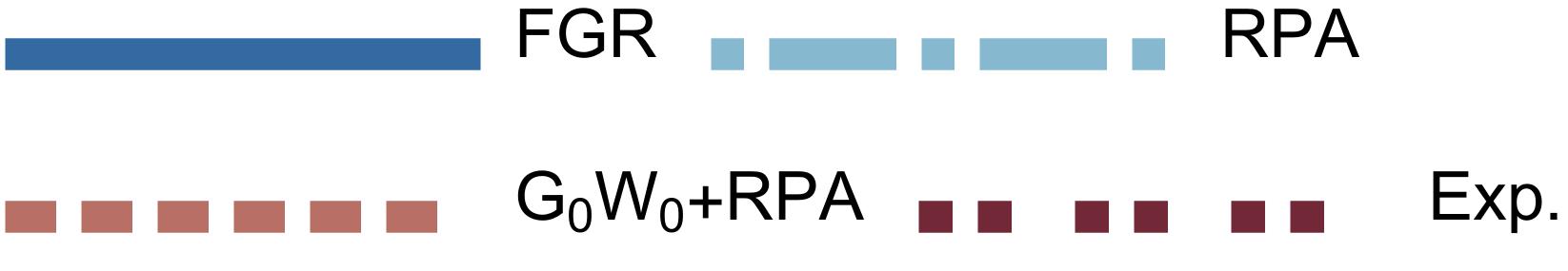}\vspace{0.25cm}

\includegraphics[width=0.3\textwidth]{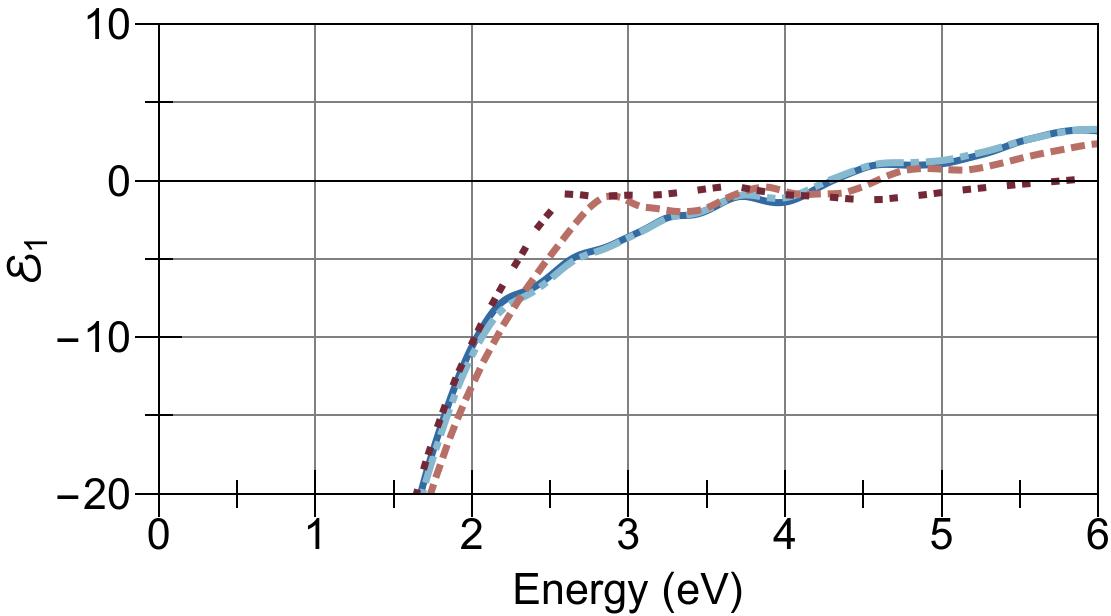}\vspace{0.25cm}

\includegraphics[width=0.3\textwidth]{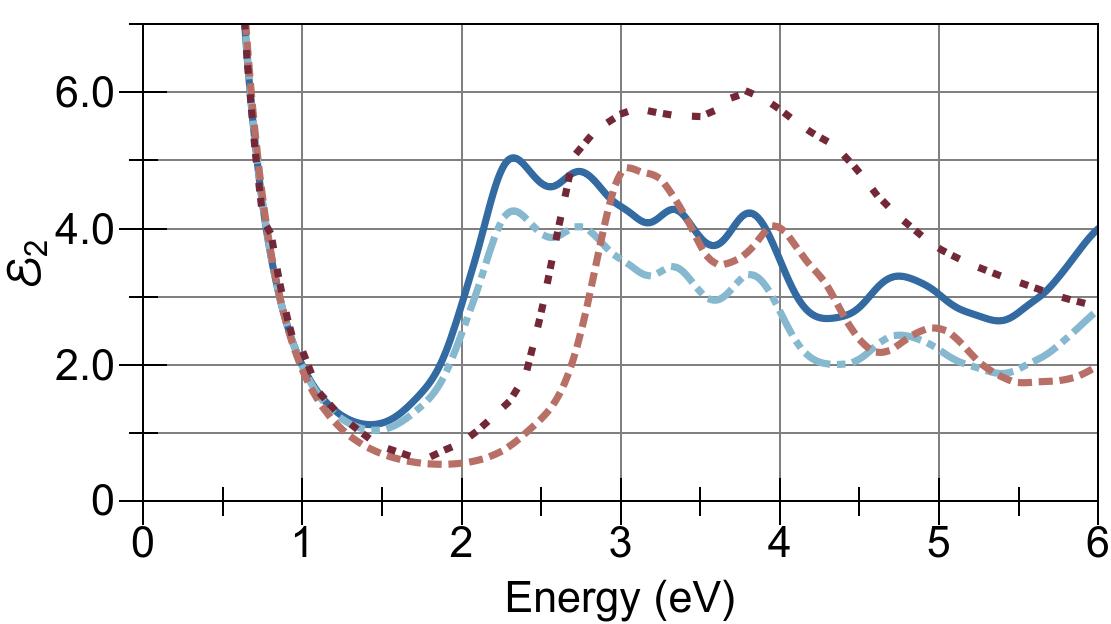}\vspace{0.25cm}

\includegraphics[width=0.3\textwidth]{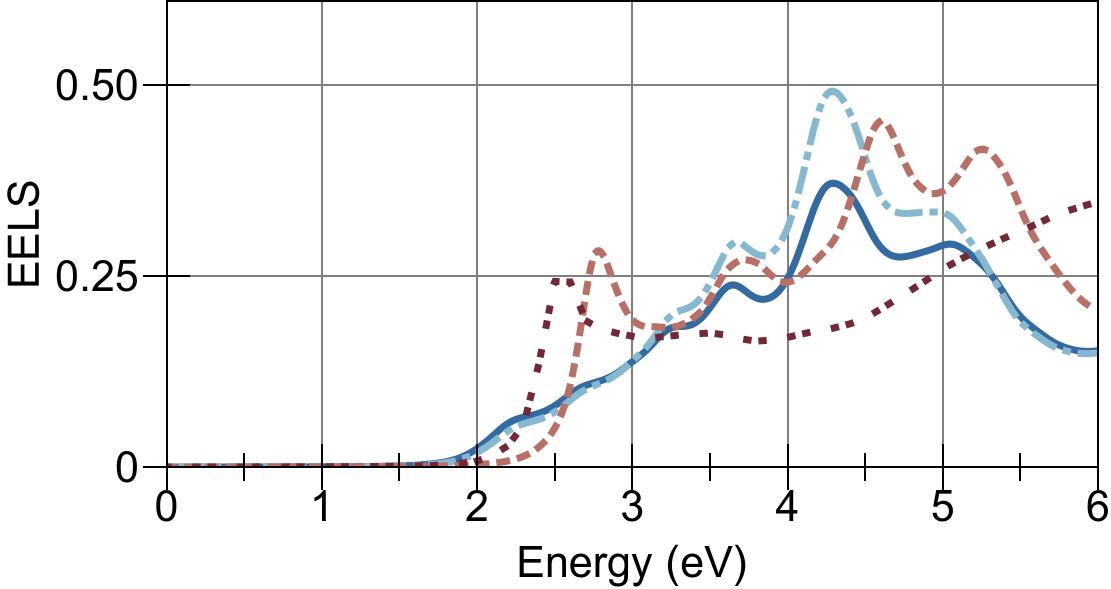}

\caption{The real and imaginary parts of the dielectric function and EELS
for bulk Au, calculated using FGR, RPA, and \gwr together with experimental data from  Babar~\cite{Babar:15}.}
\label{fig:au-spec}
\end{figure}

\begin{figure}[t]
\centering
\includegraphics[width=0.25\textwidth]{yambo-leg}\vspace{0.25cm}

\includegraphics[width=0.3\textwidth]{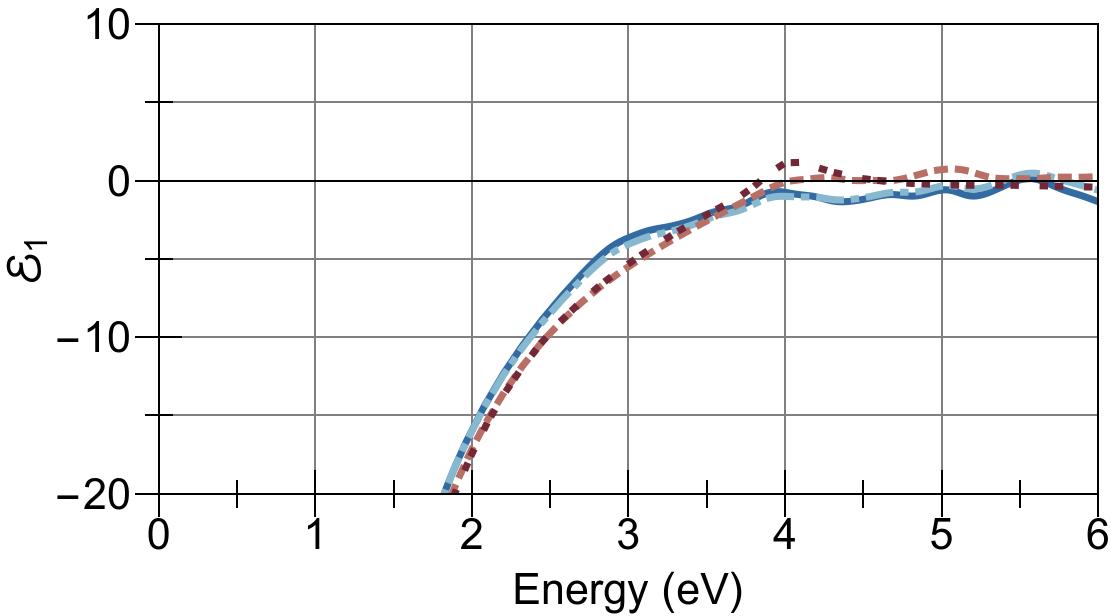}\vspace{0.25cm}

\includegraphics[width=0.3\textwidth]{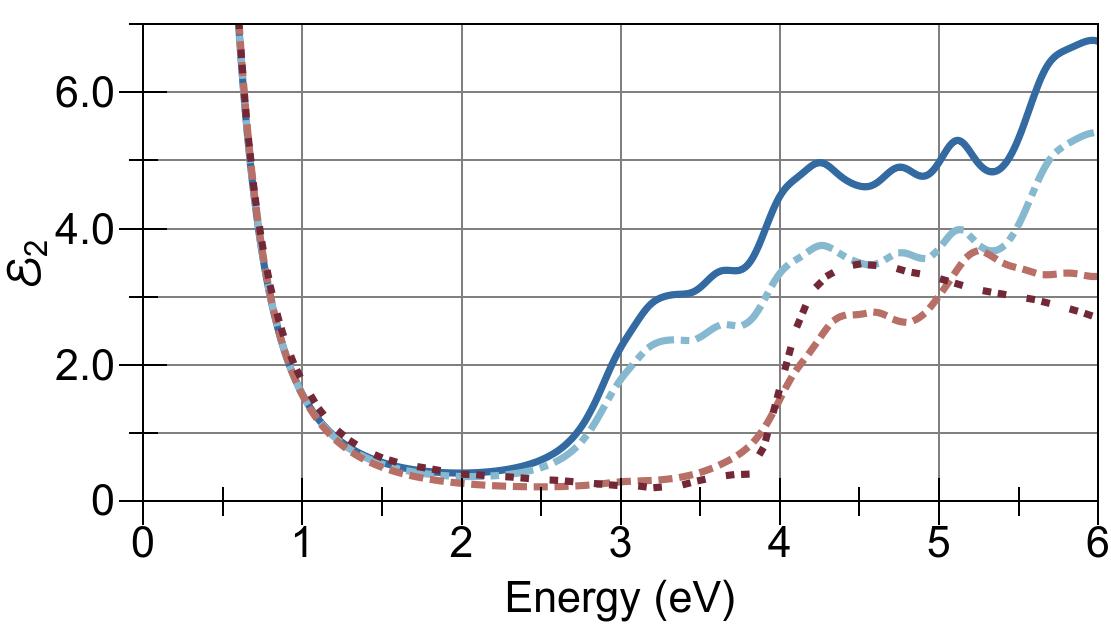}\vspace{0.25cm}

\includegraphics[width=0.3\textwidth]{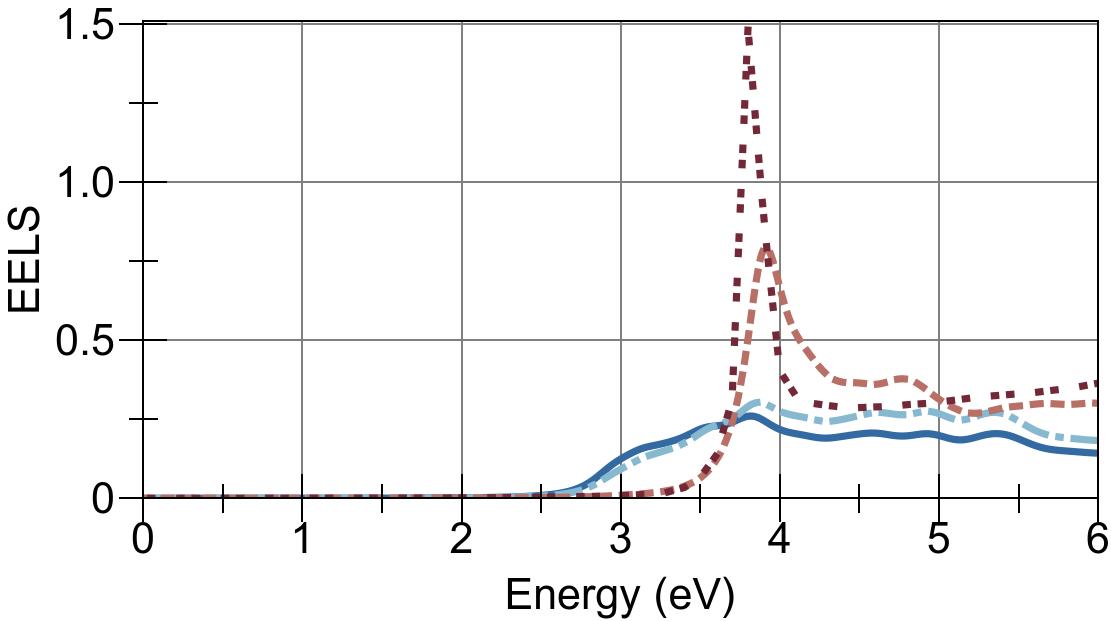}\vspace{0.25cm}
\caption{The real and imaginary parts of the dielectric function and EELS
for bulk Ag, calculated using FGR, RPA, and \gwr together with experimental data from  Babar~\cite{Babar:15}.}
\label{fig:ag-spec}
\end{figure}

\begin{figure}[t]
\centering
\includegraphics[width=0.25\textwidth]{yambo-leg}\vspace{0.25cm}

\includegraphics[width=0.3\textwidth]{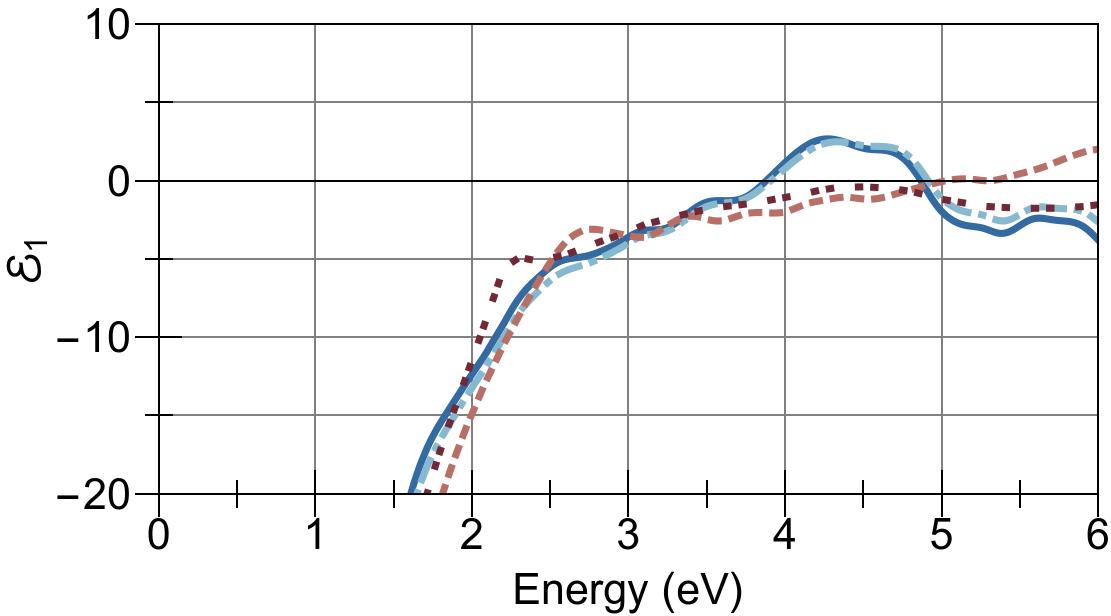}\vspace{0.25cm}

\includegraphics[width=0.3\textwidth]{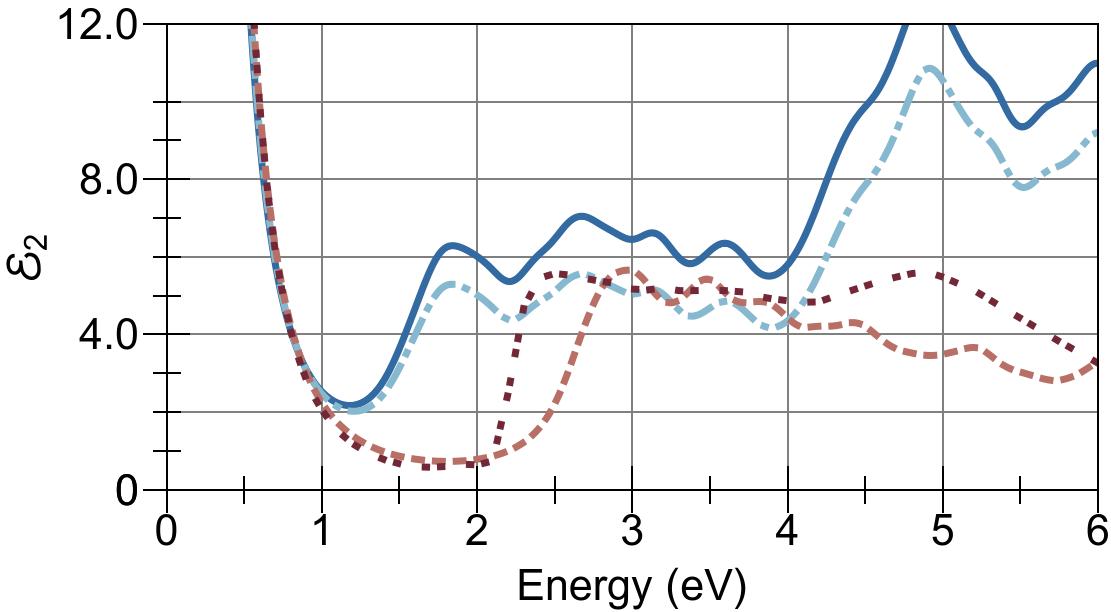}\vspace{0.25cm}

\includegraphics[width=0.3\textwidth]{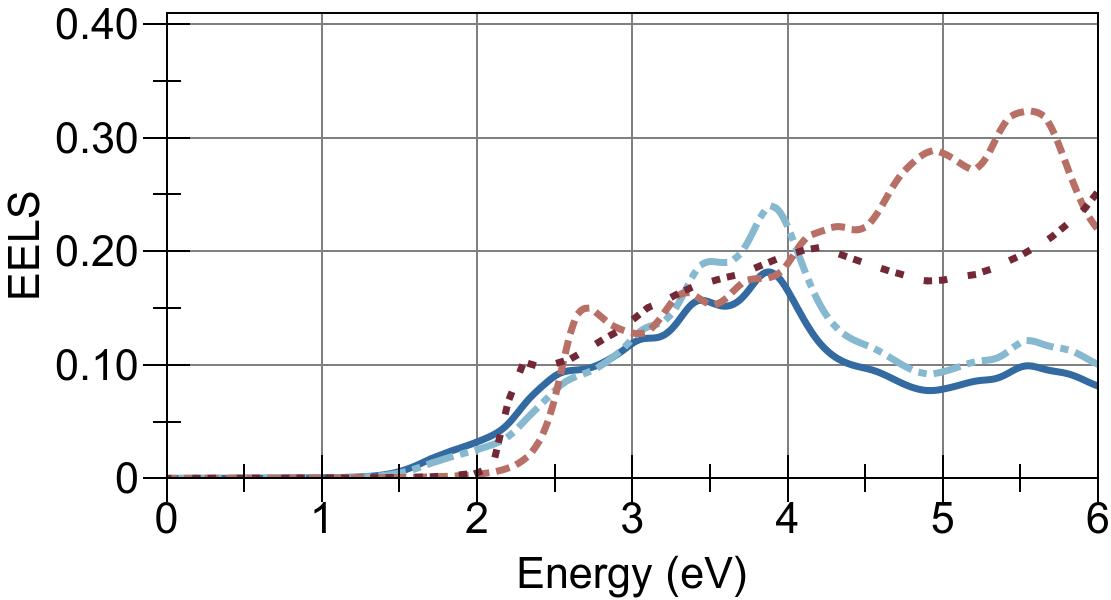}

\caption{The real and imaginary parts of the dielectric function and EELS
for bulk Cu, calculated using FGR, RPA, and \gwr together with experimental data from  Babar~\cite{Babar:15}.}
\label{fig:cu-spec}
\end{figure}

The spectra of our elemental metals were obtained by applying FGR and RPA to
the approximate KS-DFT band-structures, and RPA upon our
approximate QP band-structures.
The real and imaginary part of the total dielectric function  are shown  for pure metals in Figs.~\ref{fig:au-spec},~\ref{fig:ag-spec}, and~\ref{fig:cu-spec}, along with the experimental spectra from the detailed work by Babar and Weaver in Ref.~\citenum{Babar:15}.
Also, the electron energy-loss spectrum  (EELS)~\cite{Boersch1962,raether1965electron,Froitzheim1977,PhysRev.101.554,PhysRev.82.625,PhysRev.85.338, PhysRev.92.609} is shown, 
as calculated using the relation~\cite{sottile2003response} 
\begin{align}\label{eq:s4e28}
\mathrm{EELS}(\omega)=-\mathfrak{Im}\left[\mathcal{E}^{-1}(\omega)\right].
\end{align}

For all three systems, both FGR and RPA predict the  lowest band-to-band absorptions to be at energies $\sim$ 0.5-1.5 eV lower than those of  the experimental absorption spectra.
Furthermore, for higher energies, both approaches produce strong absorption peaks that contradict experiments.
On the other hand,  \gwr  locates the low-lying peaks at $\sim$3-4 eV in Au,  at $\sim$4-5 eV in Ag, and at $\sim$2-3 eV in Cu more accurately, and does well for the overall  curve trend,  with  respect to the experimental absorption spectra.
All three approaches reproduce the behaviour of  $\mathcal{E}_1^\textrm{Cu}$ and  $\mathcal{E}_1^\textrm{Au}$ successfully, and particularly so for $\mathcal{E}_1^\textrm{Ag}$. 
Such  improvements,  along with the very substantial improvement in $\mathcal{E}_2$ given by G\td{0}W\td{0}+RPA,  lead us to locate the first plasmonic peaks in all systems quite accurately in EELS, where FGR and RPA miss the the salient features completely (see Figs.~\ref{fig:au-spec},~\ref{fig:ag-spec}, and~\ref{fig:cu-spec}). 
As one can observe,  the improvements provided by \gwr become less effective at higher energies, although  some improvements are still achieved with respect to FGR and RPA.
This depletion of performance at higher energies is to be expected,  as the QP band-structures were approximated using an averaged stretching factor determined using only bands close to the Fermi level.
Hence, by construction, our streamlines approach is more effective for transitions between bands close to the Fermi level, which constitute the lower part of the spectra, which are always those relevant to practical plasmonic applications. 

\section{Spectra of Au$_{\textbf{x}}$Ag$_{\textbf{y}}$Cu$_{\textbf{1-x-y}}$ alloys}

Initial geometries for selected ordered alloys with compositions in multiples of $12.5\%$  were constructed by using  super cells of pure metals and substituting the desired number of atoms of other species to achieve primitive unit cells for each stoichiometric ratio.
These geometries were optimised at the DFT level.
Sample crystal structures for each stoichiometric ratio  are shown in the SI.
Alloys with the stoichiometric ratios of $\{2:1:1\}$ and $\{6:1:1\}$  have $3$ and $4$ possible phases,  respectively, and in  total $39$ structures were studied through the standardized work-flow shown in Fig.~\ref{fig:c6-wf}.
For  systems with multiple primitive phases, we calculated our final spectra by means of a thermodynamic averaging process using the Boltzmann factor defined as 
\begin{align}\label{eq:s5e29}
p_i= \exp \left( \frac{E_0-E_i}{k_B T} \right) \Bigg( \sum_j \exp \left( \frac{E_0-E_j}{k_B T} \right) \Bigg)^{-1},
\end{align}
where  $j$ is the phase index, and $E_0$ and $E_i$, respectively,  are the lowest ground-state energy among all phases and the ground-state energy of the  $i^{th}$ phase. 
The final spectra of $\{2:1:1\}$ and $\{6:1:1\}$ are thus  linear combinations of the spectra of their respective phases,  with corresponding weighting constants.

\subsection{Stretching operators for alloys}
\begin{figure}[t]
\centering
\subfloat[Z\td{v} \label{fig:alloy-v-z}]{\includegraphics[width=0.23\textwidth]{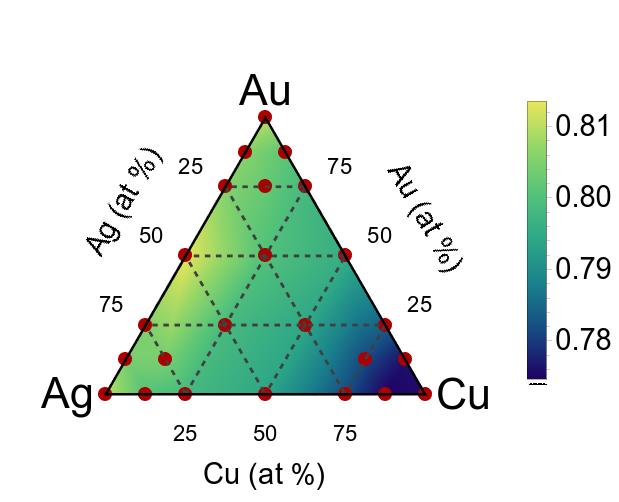}}\hfill
\subfloat[Z\td{c} \label{fig:alloy-c-z}]{\includegraphics[width=0.23\textwidth]{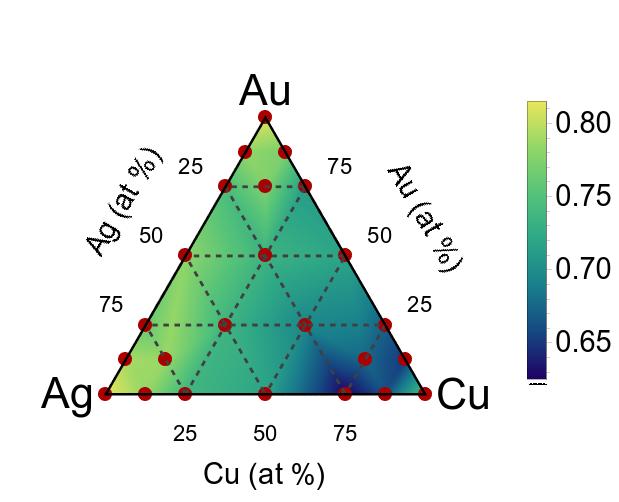}}

\subfloat[s\td{v} \label{fig:alloy-v-sciss}]{\includegraphics[width=0.23\textwidth]{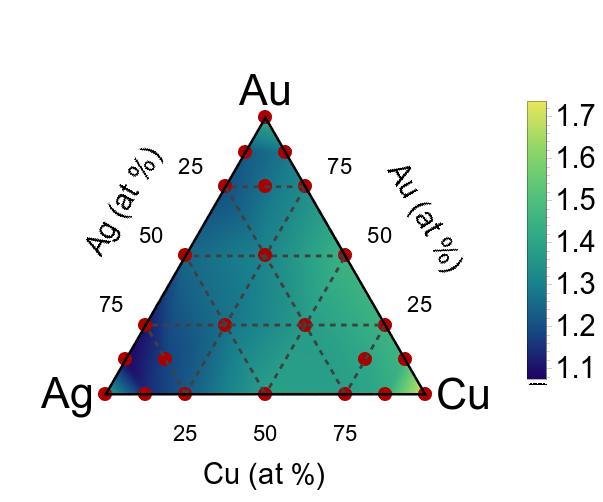}}\hfill
\subfloat[s\td{c} \label{fig:alloy-c-sciss}]{\includegraphics[width=0.23\textwidth]{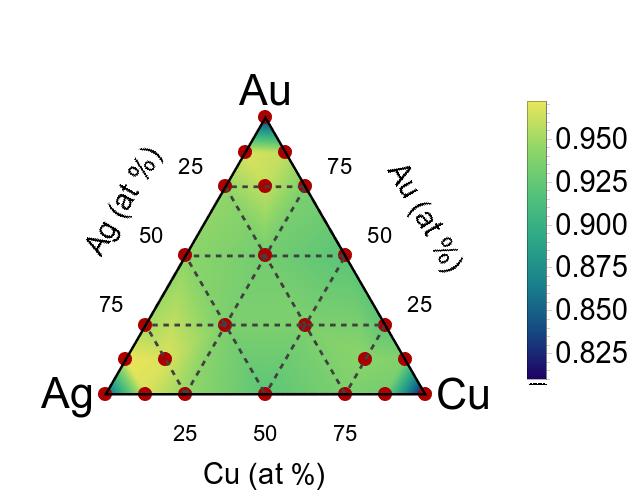}}
\caption{Stoichiometric dependence of the averaged QP renormalization factors Z (\textbf{a}, and \textbf{b}), and corresponding stretching operators (\textbf{c},and \textbf{d}) for chosen valence and conduction bands. The actual data points are marked with the red dots.}
\label{fig:alloy-gw}
\end{figure}

Band stretching operators were calculated for all valence bands and an equal number of conduction bands at 10 points at and around the $\Gamma$  point  for each crystal structure of each stoichiometric ratio.
The QP renormalization factors expressed in Eq.~\eqref{eq:s2e20} show a steady  decreasing trend with increasing Cu ratio, both in the valence and the conduction manifolds, as shown  in Fig.~\ref{fig:alloy-v-z} and Fig.~\ref{fig:alloy-c-z}.
The valence stretching operator in Fig.~\ref{fig:alloy-v-sciss} grows significantly larger for increasing Cu ratios.
Pure Cu has the largest reciprocal unit cell, where the $d$-bands are spuriously flattened and narrowly packed the most by KS-DFT,  as  seen in Fig.~\ref{fig:cu-band}.
Hence, a  larger QP stretching of the  valence manifold is expected to result from an  increasing Cu concentration. 
The conduction stretching operators show inverse trends,  providing a slightly smaller stretching at the conduction manifold for increasing Cu concentrations.
The stretching operators were applied to converged geometry of  each ordered alloy cell, and perturbative G$_0$W$_0$ simulations were performed to construct the stretched (QP) band-structures individually.
For simplicity, we will only discuss our approximated QP results from here, omitting
FGR and RPA.

\subsection{Drude parameters for alloys}
\begin{figure}[t]
\centering
\subfloat[$v_\mathrm{F}$ ($10^5$ m/s) \label{fig:alloy-vf}]{\includegraphics[width=0.23\textwidth]{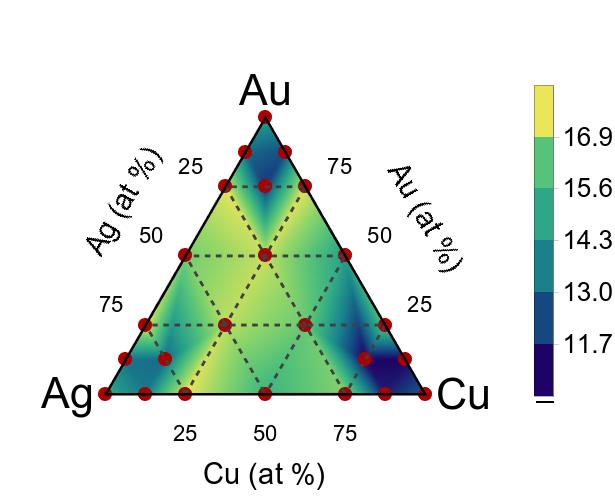}}\hfill
\subfloat[DOS at the Fermi level \label{fig:alloy-dos}]{\includegraphics[width=0.23\textwidth]{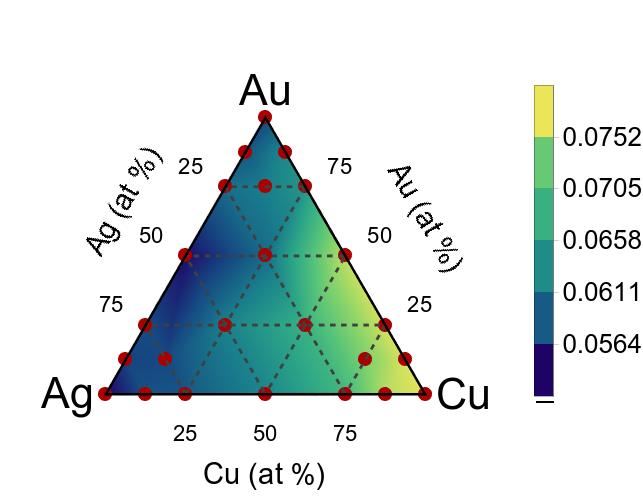}}

\caption[Stoichiometric dependence of the averaged QP Fermi velocities and DOS at the Fermi level.]{Stoichiometric dependence of the averaged QP Fermi velocities and DOS at the Fermi level (dimensionless). The actual data points are marked with the red dots.}
\label{fig:alloy-dp-pre}
\end{figure}

Using a  similar procedure to that of pure metals, the Fermi velocities and DOS at the Fermi level were computed for each alloy system.
The stoichiometric dependence of the averaged QP Fermi velocities and DOS at the Fermi level  are shown in Fig.~\ref{fig:alloy-dp-pre}. 
The Fermi velocities show some symmetric features, while they are lower in the case of predominantly Cu-based stoichiometric ratios.
Conversely, the DOS becomes large for increasing Cu ratios as   the volumes of the systems are also contracting with increasing Cu concentrations.

\begin{figure}[t]
\centering
\subfloat[$\omega_\mathrm{p}$ (eV) \label{fig:alloy-wp}]{\includegraphics[width=0.23\textwidth]{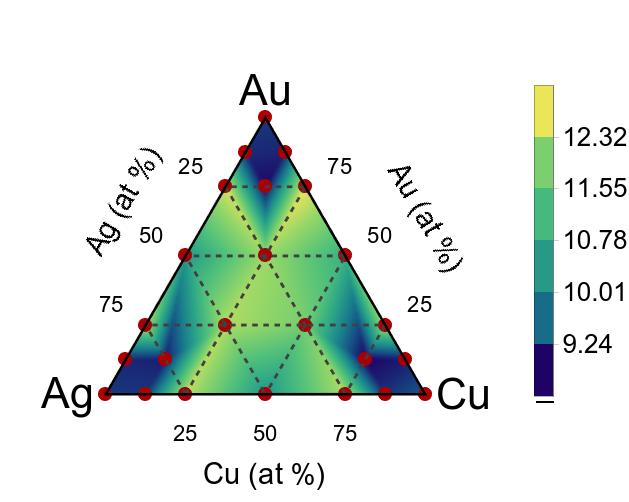}}\hfill
\subfloat[$\eta_\mathrm{p}$ (eV) \label{fig:alloy-eta}]{\includegraphics[width=0.23\textwidth]{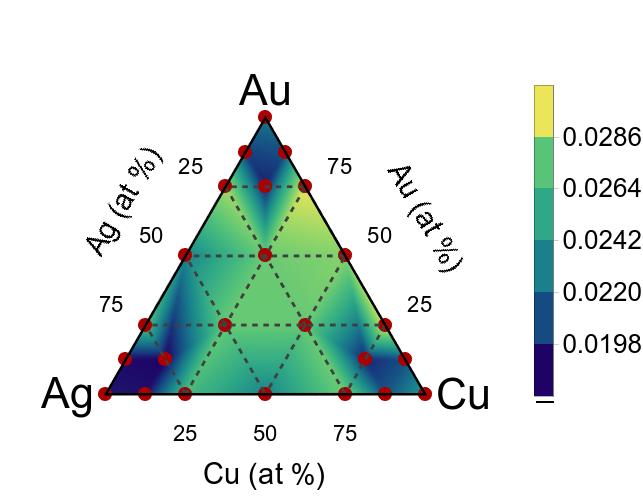}}

\subfloat[$\varepsilon_\infty$(arb.) \label{fig:alloy-epinf}]{\includegraphics[width=0.23\textwidth]{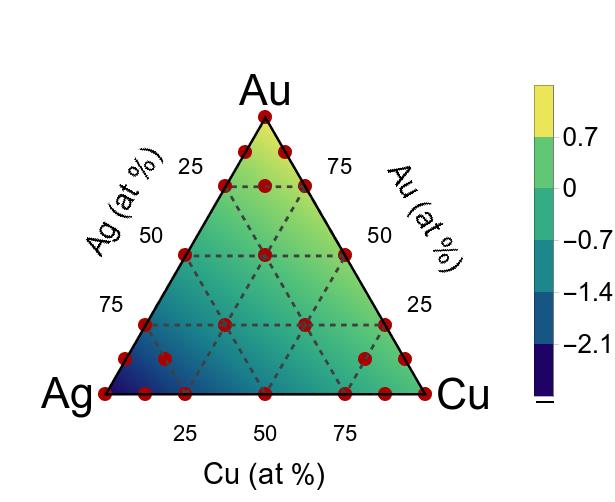}}

\caption[Stoichiometric dependence of the averaged Drude parameters for the QP band-structures.]{Stoichiometric dependence of the averaged  Drude plasmon energies, the  inverse life-times and  electric permittivity in the infinite-frequency limit  for the QP band-structures. The actual data points are marked with the red dots.}
\label{fig:alloy-dp}
\end{figure}

Competing trends in the Fermi velocities in Fig.~\ref{fig:alloy-vf} and DOS in Fig.~\ref{fig:alloy-dos} compensate each other, and lead to a symmetric trend in the  interpolated contour-plot of  the Drude plasmon energies in Fig.~\ref{fig:alloy-wp}.
The inverse life-times in Fig.~\ref{fig:alloy-eta} were evaluated by using Eq.~\eqref{eq:s2e13}, where $\{c_\eta\}$ coefficients were produced via arithmetical averaging of the inverse-life times of pure metals  with respect to the stoichiometric ratios as
\begin{align}\label{eq:s5e30}
c^\mathrm{(Au_xAg_yCu_{1-x-y})}_\eta=\left(x c_\eta^\mathrm{Au}+ y c_\eta^\mathrm{Ag}+(1-x-y)c_\eta^\mathrm{Cu}\right)
\end{align} 
As the inverse life-times are proportional to the Fermi velocities and the DOS at the Fermi level, they show similar symmetries to the Drude plasmon energies. 
The electric permittivities at the infinite-frequency limit are simply approximated by  the arithmetical averaging  of values of pure metals  with respect to the stoichiometric ratios; hence, the trend is a flat plane by construction.

\section{Plasmonic performances of  Au$_{\textbf{x}}$Ag$_{\textbf{y}}$Cu$_{\textbf{1-x-y}}$ alloys}
Even though, plasmon production is highly dependent on size and geometry of nano-materials, some fundamental criteria can be suggested as universal conditions for strong plasmonic response.  
The primary requirement of a strong plasmon is the presence of a high-density of free electrons such as in  noble and alkali metals, which are prominent systems for plasmonic applications~\cite{0953-8984-22-14-143201}. 
Moreover, plasmon quality is predominantly determined by loss~\cite{PhysRevB.67.201101,Arnold:09}, which can occur through various phenomena such as radiative dumping, surface scattering, thermal loss~\cite{Kreibig1995},  and imperfections in materials such as surface roughness~\cite{Raether1988}, and grain boundaries~\cite{doi:10.1063/1.2987458}. 
Some approximate methods are suggested to sum individual contributions of these conditions to determine overall plasmon quality~\cite{Kreibig1995}.  
As we work on perfectly ordered bulk systems, our aim is to determine some universal preliminary optical merits, which are independent of size and structural properties, to measure plasmon qualities starting from bulk dielectric functions. 
Plasmons in bulk systems are predominantly bulk plasmons, which  is a result of a combination of both intra-band and inter-band transitions~\cite{0953-8984-22-14-143201}. 
The bulk  plasmon energy $\Omega_\mathrm{p}$ is expected to be at a lower energy than the bare plasmon energy $\omega_\mathrm{p}$ due to screening of inter-band transitions. 
In addition to bulk plasmons,  there are surface plasmons due to the finite size of realistic systems. 
EELS provides a signature for bulk plasmons, whereas it requires a slight modification due to the presence of $d$-bands to capture surface plasmons~\cite{ROCCA19951}
\begin{align}\label{eq:s6e31}
-\Im{\frac{1}{1+\ep}}, 
\end{align}
which create a condition as $\ep_1\approx$ -1 and $\ep_2 \approx$ 0 for significant surface plasmons. 
Furthermore, some optical measures for  more specific plasmons such as the localized surface-plasmon (LSP)~\cite{1367-2630-4-1-393,doi:10.1021/jp0361327},   and the surface-plasmon polaritons (SPP)~\cite{ZAYATS2005131}, which are crucial to many plasmonic applications~\cite{1464-4258-5-4-353} such as optical circuits~\cite{Engheta1698} and switching~\cite{MacDonald2008,doi:10.1063/1.1650904}. 
In Ref.~\citenum{Arnold:09}, various  measures using the relation between $\ep_1$ and $\ep_2$ for LSP and SPP   at low-loss and nearly an electrostatic limit have been investigated for their respective optimized geometries. 
$\ep_1$ provides the necessary condition by checking the preliminary condition of a presence of free electrons, while $\ep_2$ is related to loss due to the decaying of plasmons to particle-hole pairs via absorption. 
Hence, the number of electrons going through inter-band transitions is desired to be small around plasmon frequencies. 
Blaber, and \textit{et al.}~\cite{0953-8984-22-14-143201} suggest some universal quality factors for LSP and SSP in metallic systems with  optimized geometries  as
\begin{align}\label{eq:s6e32}
Q_{\mathrm{LSP}}(\omega)=-\frac{\mathcal{E}_1}{\mathcal{E}_2}, \hspace{0.5cm}\mathrm{and}\hspace{0.5cm}
Q_{\mathrm{SPP}}(\omega)=\frac{\mathcal{E}_1^2}{\mathcal{E}_2}.
\end{align} 
These quality factors provide some preliminary insights on the capacity of metals to produce surface plasmons and life-times of such plasmons determined by dumping due to inter-band transitions.

\subsection{Plasmonic response at common solid-state laser wavelengths}
Three common solid-state laser wavelengths were chosen to demonstrate the plasmonic efficiencies of \alloy{x}{y}{1-x-y}. 
The first wavelength is a common  red laser at $650$ nm ($1.9074$ eV) produced using InGaAIP ~\cite{koechner2013solid} and used in a wide range of applications such as in imaging and sensing in biological systems in conjunction with a heterostructure of noble metals~\cite{doi:10.1021/ar7002804,ANIE:ANIE200804518}.
\begin{figure}[H]
\centering
\subfloat[]{\includegraphics[width=0.23\textwidth]{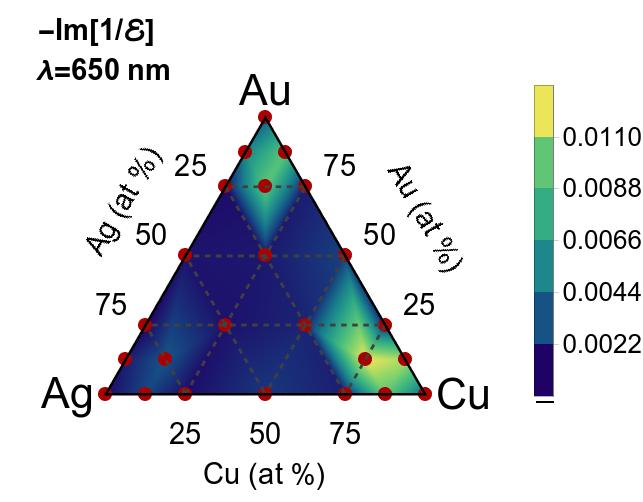}}\hfill
\subfloat[]{\includegraphics[width=0.23\textwidth]{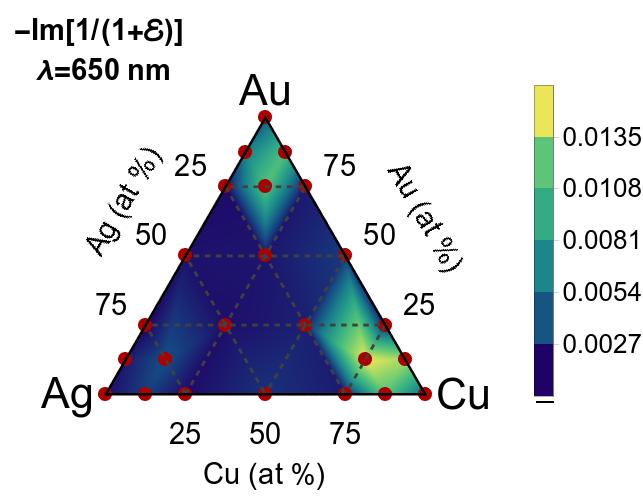}}

\subfloat[]{\includegraphics[width=0.23\textwidth]{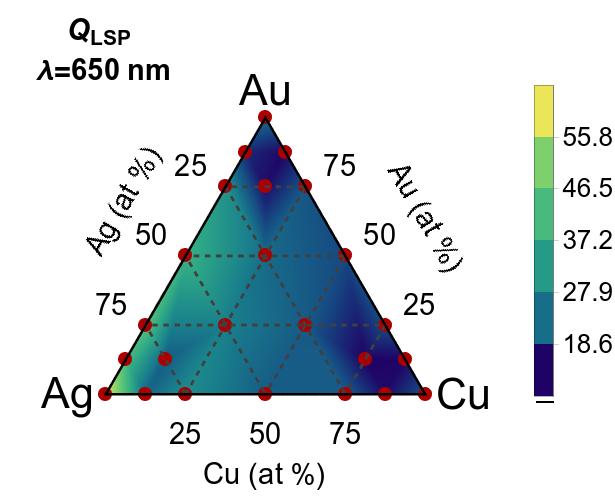}}\hfill
\subfloat[]{\includegraphics[width=0.23\textwidth]{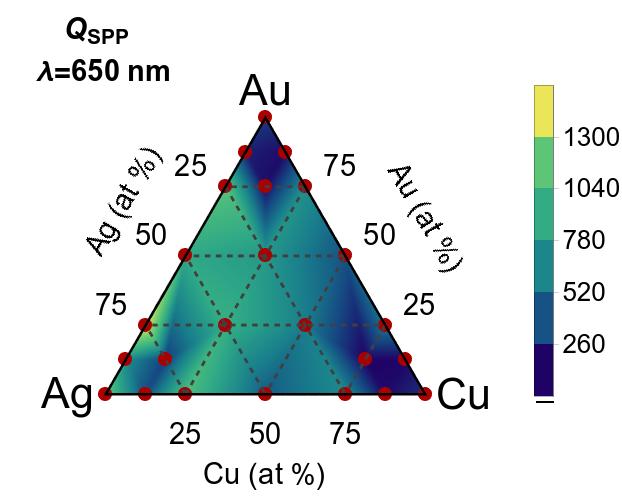}}

\caption{Stoichiometric dependence of the averaged  plasmon descriptors of bulk and surface EELS, Q\td{LSP}, and Q\td{SPP}  for  \alloy{x}{y}{1-x-y} at the common red solid-state laser wavelength 650 nm.}
\label{fig:red-desc}
\end{figure}
In Fig.~\ref{fig:red-desc}, \alloy{x}{y}{1-x-y} alloys have weak bulk and surface plasmon resonances in  orders of 10\tu{-2}- 10\tu{-3}.
As  this wavelength is commonly somewhere between Drude tails and   inter-band transition edges of $\ep_2$, these weak plasmons have long life-times due to small radiative dumping, which result in large LSP and SPP factors.
\alloy{6}{}{} and   \alloy{}{}{6} exhibit relatively strong plasmon resonance, whereas LSP and SPP have poor quality factors at these stoichiometric ratios due to larger losses at this wavelength.
Pure Ag exhibits large quality factors for LSP and SPP; however, plasmonic resonances are quite weak at 650 nm.

\begin{figure}[H]
\centering
\subfloat[]{\includegraphics[width=0.23\textwidth]{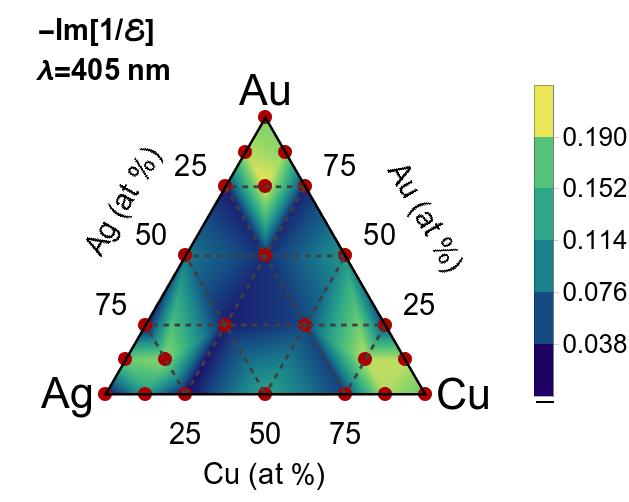}}\hfill
\subfloat[]{\includegraphics[width=0.23\textwidth]{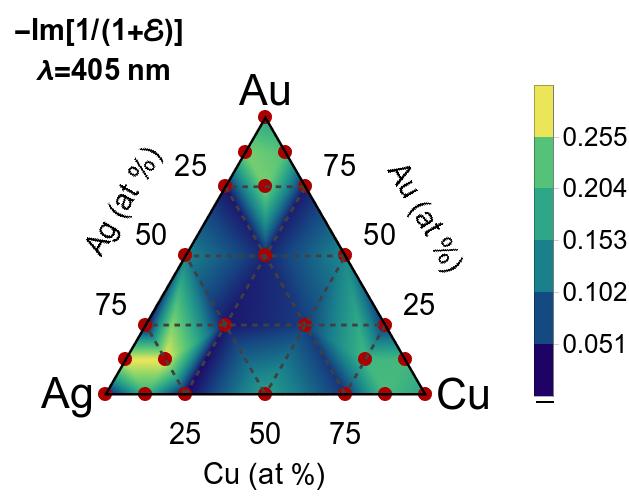}}

\subfloat[]{\includegraphics[width=0.23\textwidth]{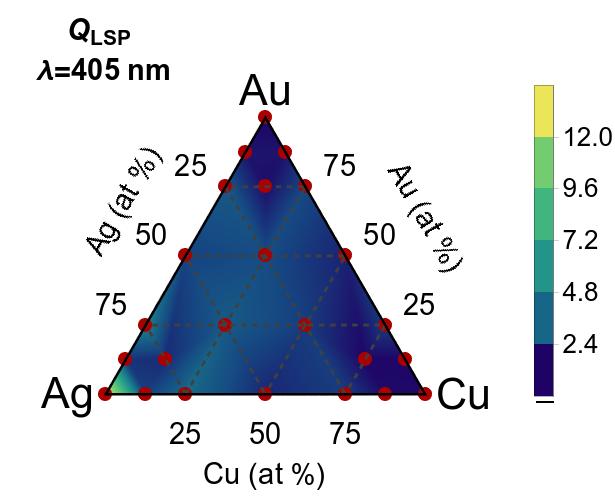}}\hfill
\subfloat[]{\includegraphics[width=0.23\textwidth]{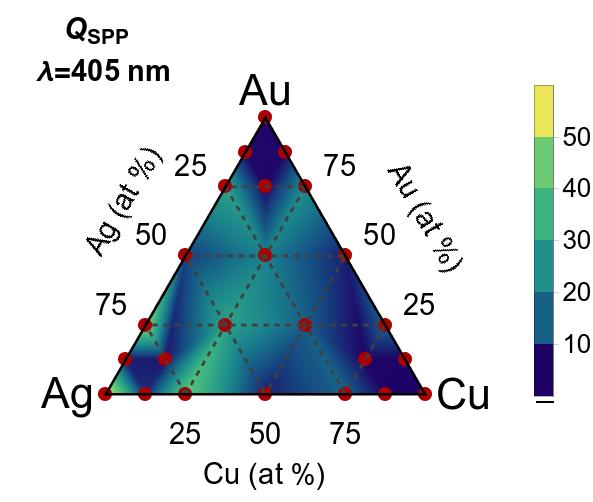}}

\caption{Stoichiometric dependence of the averaged plasmon descriptors of bulk and surface EELS, Q\td{LSP}, and Q\td{SPP} for  \alloy{x}{y}{1-x-y} at the common blu-ray solid-state laser wavelength 405 nm.}
\label{fig:blu-desc}
\end{figure}

\begin{figure}[H]
\centering
\subfloat[]{\includegraphics[width=0.23\textwidth]{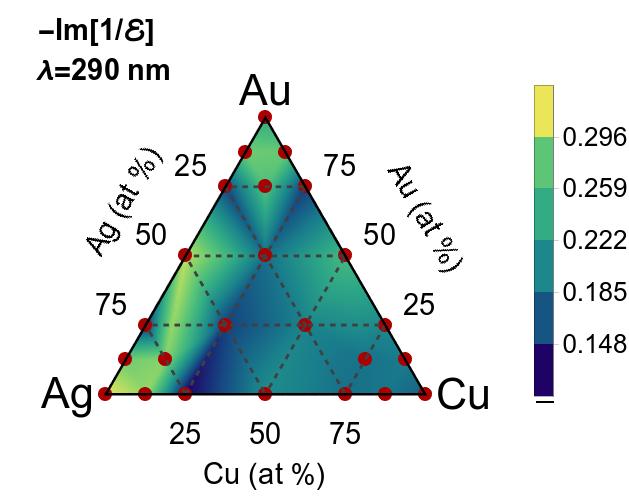}}\hfill
\subfloat[]{\includegraphics[width=0.23\textwidth]{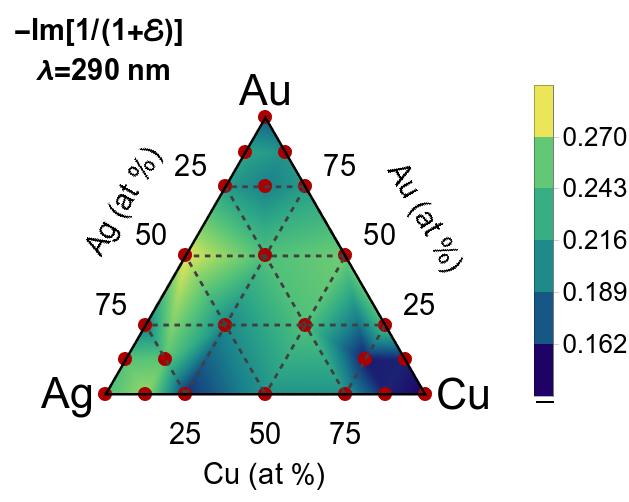}}

\subfloat[]{\includegraphics[width=0.23\textwidth]{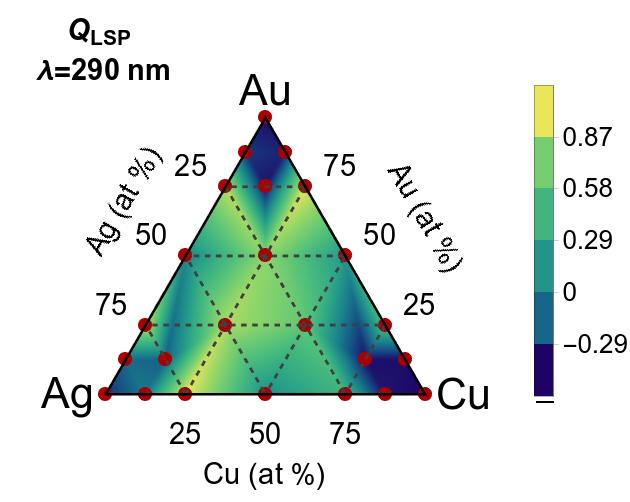}}\hfill
\subfloat[]{\includegraphics[width=0.23\textwidth]{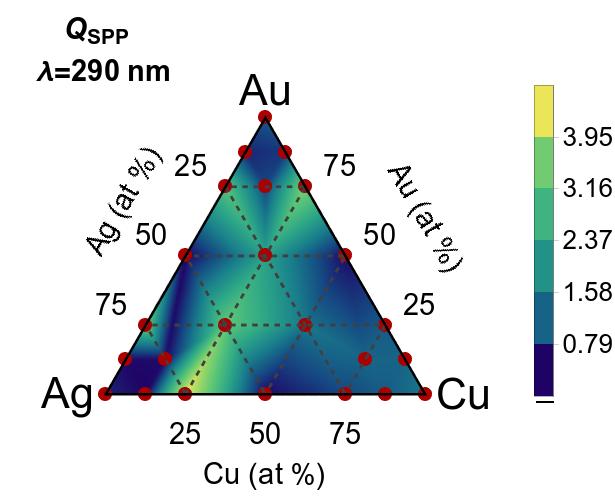}}

\caption[Stoichiometric dependence of the averaged bulk and surface EELS, Q\td{LSP}, and Q\td{SPP} for  \alloy{x}{y}{1-x-y} at the common deep-UV solid-state laser wavelength 290nm.]{Stoichiometric dependence of the averaged plasmon descriptors of bulk and surface EELS, Q\td{LSP}, and Q\td{SPP} for  \alloy{x}{y}{1-x-y} at the common deep-UV solid-state laser wavelength 290nm.}
\label{fig:duv-desc}
\end{figure}

The second frequency is that of the  common lasers used in blu-ray devices at $405$ nm ($3.0613$ eV),  generally produced using InGaN~\cite{1347-4065-35-1B-L74} for efficient optical recording in conjunction with noble metal nano-clusters~\cite{ADMA:ADMA201002413}.  
In Fig.~\ref{fig:blu-desc}, \{7:1:0\} stoichiometric ratios show relatively significant plasmon resonance alongside  \{6:1:1\} stoichiometric ratios in order of $\sim$ 10\tu{-1}, which are comparable to these of maximums of plasmon peaks of binary alloys,  with large quality factors of LSP  and SPP.
Ag predominantly has long-lived LSP, while AuAg\td{3} shows some longevity.
Lastly, the deep-UV laser at $290$ nm ($4.2753$ eV) commonly produced in Ce:LiSAF / Ce:LiCAF media with Nd:YAG lasers~\cite{Marshall:94} was presented for the sake of discussion.
In Fig.~\ref{fig:duv-desc}, the bulk and surface EELS profiles here  have structures but  still a stronger response in Ag, Au, Au\td{6}AgCu, and AuAg\td{6}Cu,  with a newly arising strong response in AuAg as shown.  
Correspondingly, AuAg has short-lived LSP and SPP as does Ag, which performs relatively better in the former case as at the blu-ray laser.

The plasmonic responses of \alloy{x}{y}{1-x-y} have complex behaviours, which cannot be simply interpolated based on the stoichiometric ratios alone, despite the similar electronic structures of  constituting atoms.
Consistent first-principles simulations  are thus essential to engineer and predict comprehensive alloy systems with well-controlled approximations at reasonable computational costs.

\section{Conclusion}

It was shown that RPA starting from the KS band-structure fails to acceptably locate peaks in the absorption spectra of Au, Ag, and Cu,  whereas RPA starting even from  an approximate QP band-structure performs drastically better by locating low-energy peaks accurately in absorption spectra.
Such an approximate QP band-structure can be achieved with little additional computational costs by applying some average stretching to bands via stretching operators, which can be obtained within \gw for a small set of grid points in the Brillouin zone and bands around the Fermi level.
Despite their lack of  finite size and surface effects, the bulk dielectric functions were used to determine some preliminary optic merits for LSP and SPP at plasmon resonance wavelengths. 
At the common solid-state red laser around 650 nm, AuAgCu\td{6}, Au\td{6}AgCu, AuCu\td{7}, and Au\td{7}Cu show relatively strong plasmon resonances, whereas higher Ag concentrations reduce plasmon resonance in general. 
AuAg\td{6}Cu and AuAu\td{7} as well as AuAg, AuCu, and AgCu alloys start showing   stronger plasmon resonances with significant quality factors for LSP and SPP at the common blu-ray laser at 405 nm. 
This trend at 405 nm becomes more distinctive at deep-UV laser wavelengths around 290 nm. 
Particularly, pure Ag, AuAg\td{7}, Ag\td{7}Cu produce significant plasmon resonances with high LSP and SPP quality factors  at 290 nm. 
Combining RPA starting an approximate QP band-structure with the Drude-Lorentz model using the semi-classical Drude parameters provide a computationally feasible approach to investigate spectra and plasmonic responses of the noble metals and their alloys. 
Despite its simplicity, some preliminary optic merits can be obtained that are useful as a starting point for tailoring plasmonic responses in such systems. 

We gratefully acknowledge the support of 
Trinity College Dublin's Studentship Award
and  School of Physics.
We acknowledge and thank Tonatiuh Rangel
and Daniele Varsano for discussions.
We also acknowledge the DJEI/DES/SFI/HEA Irish Centre for High-End Computing (ICHEC) for the provision of computational facilities and support.
We finally acknowledge the Trinity Centre for High Performance Computing and Science Foundation Ireland 
for the maintenance and
funding, respectively, of the Boyle cluster on which
further calculations were performed.

\end{document}

\title{{\Huge Supporting Information} \\  \vspace{2cm}Plasmonic performances of   Au$_\mathbf{x}$Ag$_\mathbf{y}$Cu$_\mathbf{1-x-y}$  alloys using many-body perturbation theory}% Force line breaks with \\
%\thanks{A footnote to the article title}%

\author{Okan K. Orhan, and David D.  O'Regan}

\maketitle

\tableofcontents

\newpage 

\section{Sample crystal structures for the studied stoichiometric ratios}

\begin{figure}[H]
\centering
\begin{subfigure}[b]{0.2\textwidth}
\centering
\includegraphics[width=1.0\textwidth]{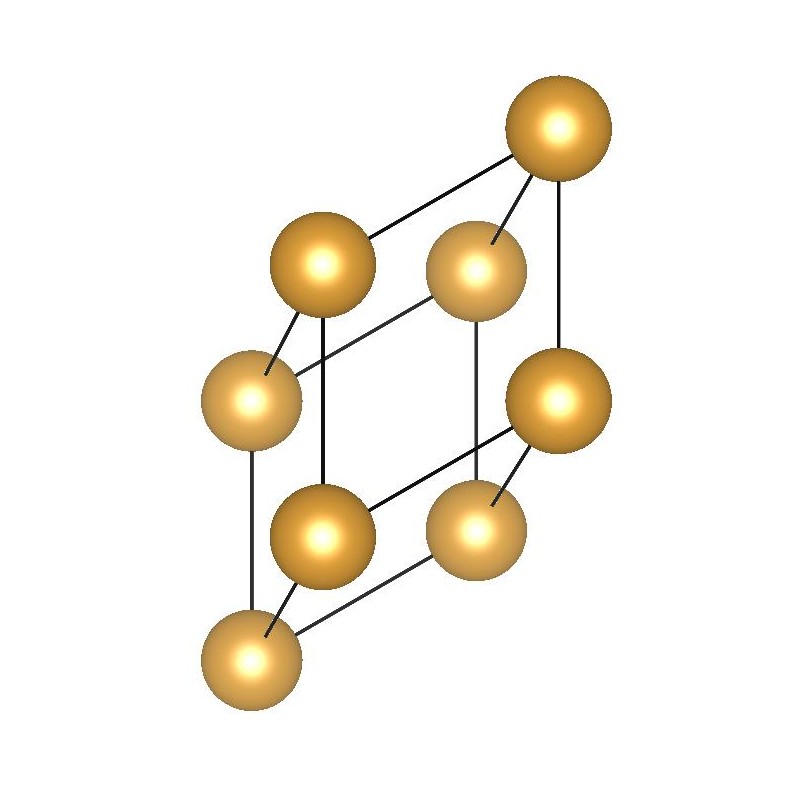}
\caption{Au}
\label{fig:au-str}
\end{subfigure}\hspace{0.5cm}
\begin{subfigure}[b]{0.2\textwidth}
\centering
\includegraphics[width=1.0\textwidth]{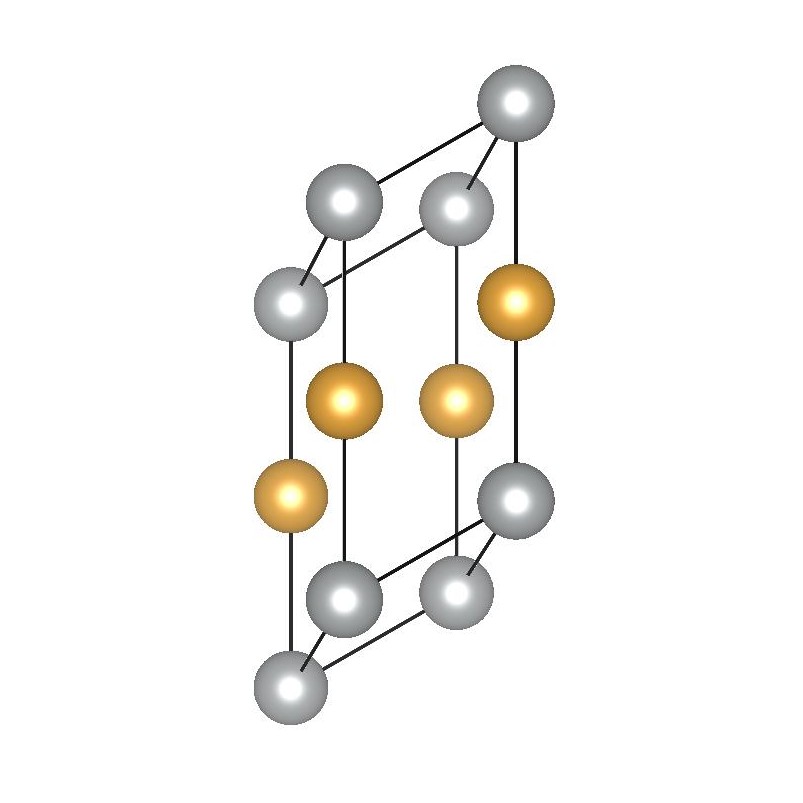}
\caption{AuAg}
\label{fig:ag-au-str}
\end{subfigure}\hspace{0.5cm}
\centering
\begin{subfigure}[b]{0.2\textwidth}
\centering
\includegraphics[width=1.0\textwidth]{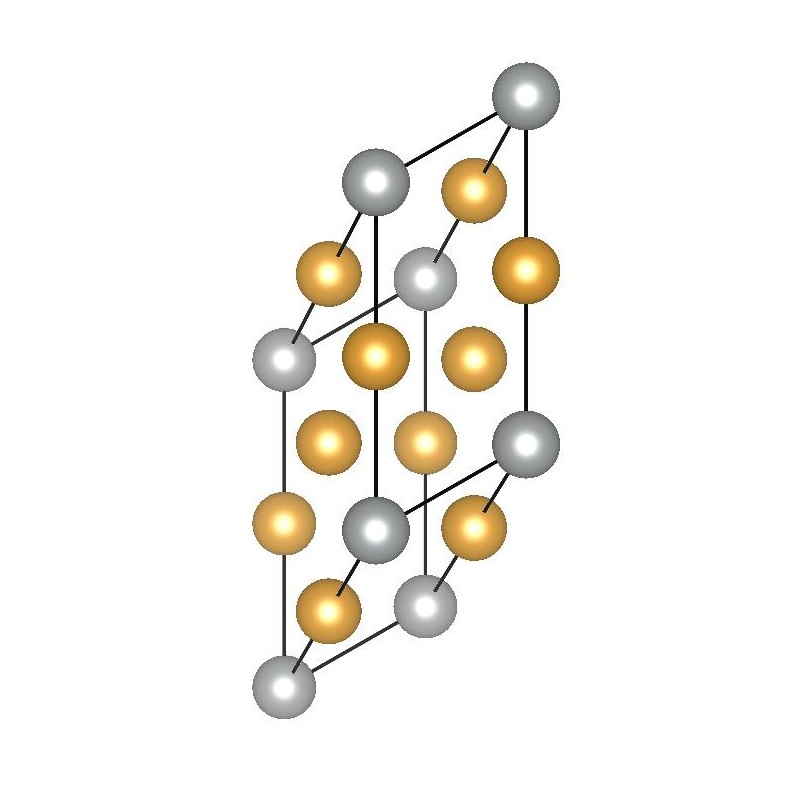}
\caption{Au\td{3}Ag}
\label{fig:ag3-au-str}
\end{subfigure}

\begin{subfigure}[b]{0.2\textwidth}
\centering
\includegraphics[width=1.0\textwidth]{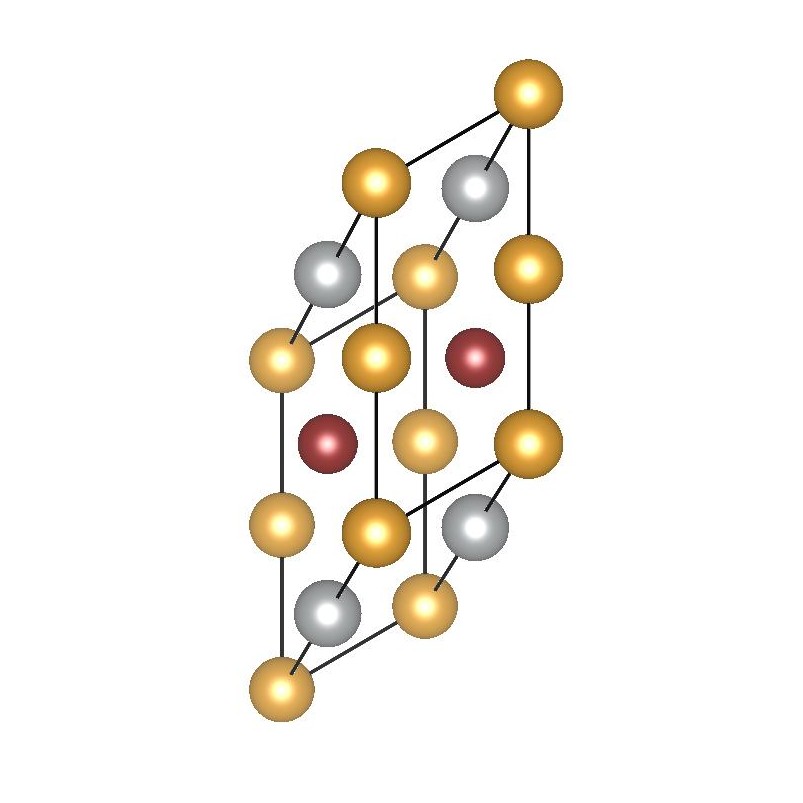}
\caption{(Au\td{2}AgCu)\td{Phase 1}}
\label{fig:ag-au2-cu-str1}
\end{subfigure}\hspace{0.5cm}
\centering
\begin{subfigure}[b]{0.2\textwidth}
\centering
\includegraphics[width=1.0\textwidth]{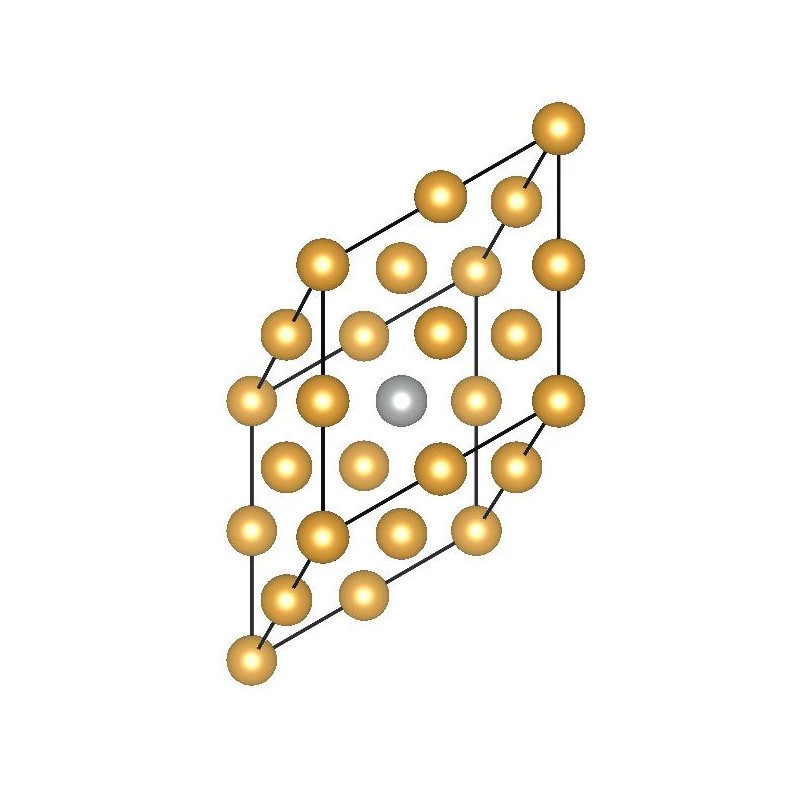}
\caption{Au\td{7}Ag}
\label{fig:ag-au7-str}
\end{subfigure}\hspace{0.5cm}
\centering
\begin{subfigure}[b]{0.2\textwidth}
\centering
\includegraphics[width=1.0\textwidth]{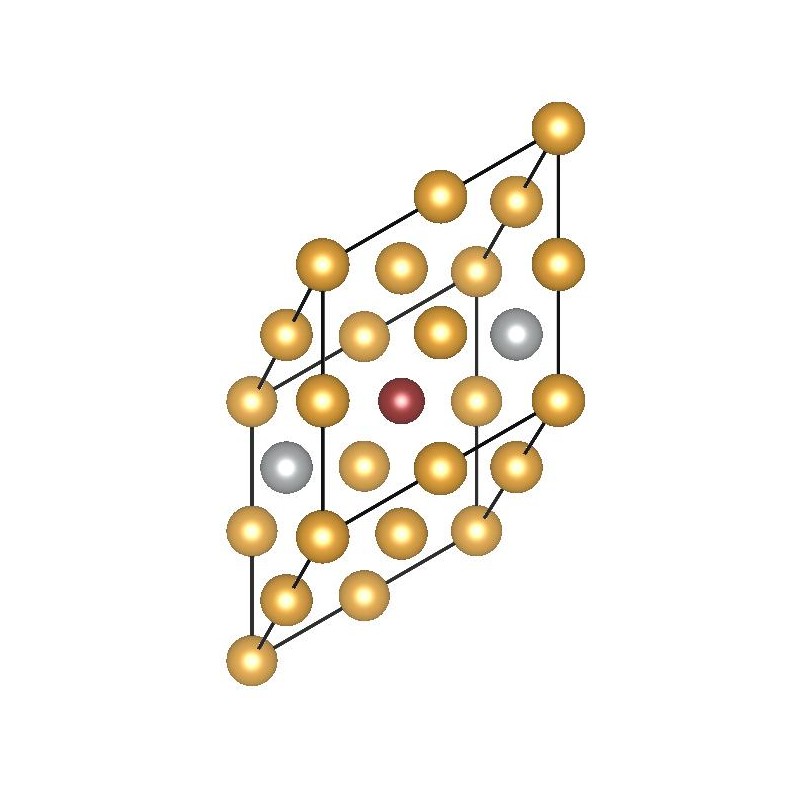}
\caption{(Au\td{6}AgCu)\td{Phase 1}}
\label{fig:ag-au6-cu-str1}
\end{subfigure}

\caption[Sample crystal structures for chosen  stoichiometric ratios of alloys.]{The sample structure for chosen stoichiometric ratios of \alloy{x}{y}{1-x-y}.}
\label{fig:alloy-str}
\end{figure}

\section{Drude Parameters}

\begin{table}[H]
\renewcommand{\arraystretch}{1.8} \setlength{\tabcolsep}{8pt}
\begin{center}
{\small
\begin{tabular}{lc:cc:cc|ccc} \hline \hline 
 & $\omega_\mathrm{p}$  & $\eta_\mathrm{p}^{\mathrm{FGR}}$  & $\varepsilon_\infty^{\mathrm{FGR}}$ & $\eta_\mathrm{p}^{\mathrm{RPA}}$  & $\varepsilon_\infty^{\mathrm{RPA}}$  & $\omega_p^{\mathrm{G_0W_0}}$  & $\eta_\mathrm{p}^{\mathrm{G_0W_0}}$  & $\varepsilon_\infty^{\mathrm{G_0W_0}}$ \\ \hline
 Au &  8.911     & 0.01990  &  2.143 & 0.01991 &  3.775  &   8.947    & 0.02266     & 1.601 \\
Ag &   8.725     & 0.01896  &  -4.436 & 0.01896  &  -2.610   &  8.925    & 0.01918  &   -2.624\\
Cu &   8.836     & 0.02756 &  -2.988 &  0.02758 &   -0.688  &  9.313     &  0.02426    & 0.104 \\
 \hline \hline
\end{tabular}}
\end{center}
\caption[The Drude parameters for pure Au, Ag, and Cu.]{The Drude parameters $\{\omega_\mathrm{p},\eta_\mathrm{p},\varepsilon_\infty\}$ (in eV, eV and dimensionless, respectively) for spectra using FGR, RPA, and \gwr for pure Au, Ag, and Cu.}
\label{tab:c6t5}
\end{table}

\begin{table}[H]
\renewcommand{\arraystretch}{1.8} \setlength{\tabcolsep}{7pt}
\begin{center}
{\small
\begin{tabular}{ lccc:lccc}
\hline \hline
 System  &  $\omega_p$  &  $\eta$  &  $\varepsilon_\infty$  &  System  &  $\omega_p$  &  $\eta$  &  $\varepsilon_\infty$
  \\ 
\hline 
Au  & 8.947 & 0.02266 & 1.601 &    Au$_7$Ag  & 8.793 & 0.01992 & 1.073\\

Ag & 8.925 & 0.01918 & -2.624 &  Au$_7$Cu  & 8.708 & 0.01991 & 1.414   \\

Cu  &  9.313 & 0.02426 & 0.1039 &     AuAg$_7$  & 8.735 & 0.01748 &  -2.096 \\

AuAg  &  10.36 & 0.02392 & -0.5116 &   AuCu$_7$  &  8.61 & 0.02041 & 0.291 \\

AuCu  &  10.91 & 0.0252 & 0.8523 &    Ag$_7$Cu  &  8.918 & 0.01835 & -2.283 \\

AgCu  & 10.32 & 0.02466 & -1.26 &   AgCu$_7$  &  8.896 & 0.02108 & -0.2371 \\

Au$_3$Ag   & 12.8 & 0.0278 & 0.5446 &  Au$_6$AgCu (p1)  &  8.666 & 0.01969 & 0.8855\\

Au$_3$Cu  &  12.99 & 0.03062 & 1.227 &   Au$_6$AgCu (p2)  & 8.655 & 0.01963 & 0.8855 \\

AuAg$_3$   &  12.45 & 0.02519 & -1.568 &   Au$_6$AgCu (p3)  & 8.634 & 0.01952 & 0.8855\\

AuCu$_3$  & 12.41 & 0.02934 & 0.4781 &    Au$_6$AgCu (p4)  &  8.087 & 0.01788 & 0.8855 \\

 Ag$_3$Cu   &  12.58 & 0.02697 & -1.942 &  AuAg$_6$Cu (p1)  & 8.571 & 0.01822 & -1.755 \\ 

AgCu$_3$  &  12.51 & 0.02997 & -0.5781 &  AuAg$_6$Cu (p2)  &  8.585 & 0.01828 & -1.755 \\

Au$_2$AgCu (p1)  & 12.64 & 0.02888 & 0.1703 &   AuAg$_6$Cu (p3)  &  8.614 & 0.01838 & -1.755 \\

Au$_2$AgCu (p2)  &  12.26 & 0.02872 & 0.1703 &  AuAg$_6$Cu (p4)  & 8.896 & 0.0184 & -1.755   \\

 Au$_2$AgCu (p3)  & 12.26 & 0.02872 & 0.1703 &   AuAgCu$_6$ (p1)  & 8.437 & 0.01975 & -0.05001\\
 
 AuAg$_2$Cu (p1)  &  11.97 & 0.02676 & -0.8858 &  AuAgCu$_6$ (p2)  &  8.445 & 0.01979 & -0.05001 \\

AuAg$_2$Cu (p2)  & 12.93 & 0.02774 & -0.8858 &   AuAgCu$_6$ (p3)  &  8.338 & 0.01949 & -0.05001\\

AuAg$_2$Cu (p3)  & 12.92 & 0.02773 & -0.8858 &  AuAgCu$_6$ (p4)  &  8.246 & 0.01945 & -0.05001 \\
 
AuAgCu$_2$ (p1)  & 12.84 & 0.02931 & -0.2039 &    &   &   & \\

AuAgCu$_2$ (p2)  & 11.72 & 0.02738 & -0.2039 &    &   &   & \\

AuAgCu$_2$ (p3)  &  11.73 & 0.02739 & -0.2039 &   &   &   & \\

\hline \hline
\end{tabular}}
\end{center}
\caption{Drude parameters for \alloy{x}{y}{1-x-y}.}
\label{tab:alloy-full-dp}
\end{table}

\section{Stretching operators}

\begin{table}[H]
\renewcommand{\arraystretch}{1.8} \setlength{\tabcolsep}{10pt}
\begin{center}
{\small
\begin{tabular}{ lcc:lcc}
\hline \hline
 System  &  $s_c$  &  $s_v$  &  System  &  $s_c$  &  $s_v$
  \\ 
\hline 
Au  &  0.825253   &  1.419797  &   Au$_7$Ag  &  0.967842  &  1.218996\\

Ag  &  0.846172   & 1.376302   & Au$_7$Cu  &   0.959605  &  1.277749 \\

Cu  &  0.809883   & 1.735804  &     AuAg$_7$  &  0.972795  &  1.072632  \\

AuAg  &  0.942895  &  1.245925 &   AuCu$_7$  &  0.937301  &  1.459205  \\

AuCu  &   0.921209  &  1.451945   &  Ag$_7$Cu  &  0.966574  &  1.138571  \\

AgCu  &  0.921900  &  1.382942  &   AgCu$_7$  &  0.932309  &  1.414183  \\

Au$_3$Ag   &  0.945233  &  1.197529  &   Au$_6$AgCu (p1)  &  0.961430  &  1.272030\\

Au$_3$Cu  &   0.936102  &  1.329064  &   Au$_6$AgCu (p2)  &  0.961097  &  1.26530\\

AuAg$_3$   &  0.951899  &  1.105431   &  Au$_6$AgCu (p3)  &  0.961277  &  1.266095 \\

AuCu$_3$  &  0.922466  &  1.454606  &   Au$_6$AgCu (p4)  &  0.959646  &  1.268756\\

 Ag$_3$Cu   &  0.945817  &  1.215392  &  AuAg$_6$Cu (p1)  &  0.965500  &  1.168711 \\ 

AgCu$_3$  &  0.935669  &  1.382684  &   AuAg$_6$Cu (p2)  &  0.965551  &  1.177290 \\

Au$_2$AgCu (p1)  & 0.936355   & 1.302908    &  AuAg$_6$Cu (p3)  &  0.974523  &  1.185852 \\

Au$_2$AgCu (p2)  &  0.931013  &  1.311556  &  AuAg$_6$Cu (p4)  &  0.965523  &  1.174112  \\

 Au$_2$AgCu (p3)  & 0.930925  &  1.311619  &  AuAgCu$_6$ (p1)  &  0.936233  &  1.404183 \\
 
 AuAg$_2$Cu (p1)  &  0.938311  &  1.272339  &   AuAgCu$_6$ (p2)  &  0.943095  &  1.399728  \\

AuAg$_2$Cu (p2)  &   0.934446  &  1.261742    &    AuAgCu$_6$ (p3)  &  0.936821  &  1.403408\\

AuAg$_2$Cu (p3)  &  0.934076   & 1.261342  &   AuAgCu$_6$ (p4)  &  0.943707  &  1.407826 \\
 
AuAgCu$_2$ (p1)  &  0.933284  &  1.382637   &    &   & \\

AuAgCu$_2$ (p2)  &  0.936700  &  1.371520   &     &   & \\

AuAgCu$_2$ (p3)  &  0.936793  &  1.372440  &     &   & \\

\hline \hline
\end{tabular}}
\end{center}
\caption{First-principles \gw stretching operators for conduction and valance bands of \alloy{x}{y}{1-x-y}.}
\label{tab:alloy-full-sciss}
\end{table}